\documentclass[12pt]{article}

\pdfoutput=1

\usepackage{graphics}
\usepackage{amssymb}
\usepackage{amsmath}
\usepackage{tikz}
\usepackage{subcaption}
\usepackage{enumitem}
\setlist{itemsep=1pt,topsep=2pt, leftmargin=*}

\newcommand{\beq}{\begin{equation}}

\newcommand{\eeq}{\end{equation}}
\newcommand{\bea}{\begin{eqnarray}}
\newcommand{\eea}{\end{eqnarray}}

\newcommand{\R}{\mathbb{R}}

\begin{document}

\begin{center}
${}$\\
\vspace{100pt}
{ \Large \bf Curvature profiles for quantum gravity 
}

\vspace{46pt}

{\sl J. Brunekreef}
and {\sl R. Loll}

\vspace{24pt}
{\footnotesize

Institute for Mathematics, Astrophysics and Particle Physics, Radboud University \\ 
Heyendaalseweg 135, 6525 AJ Nijmegen, The Netherlands.\\ 
\vspace{12pt}
{email: j.brunekreef@science.ru.nl, r.loll@science.ru.nl}\\
}
\vspace{48pt}

\end{center}

\vspace{0.8cm}

\begin{center}
{\bf Abstract}
\end{center}

\noindent Building on the recently introduced notion of quantum Ricci curvature and motivated by considerations in nonperturbative
quantum gravity, we advocate a new, global observable for curved metric spaces, the {\it curvature profile}. 
It is obtained by integrating the scale-dependent, quasi-local quantum Ricci curvature, and therefore also depends
on a coarse-graining scale. 
To understand how the distribution of local, Gaussian curvature is reflected in the curvature profile, we compute it on a class of
regular polygons with isolated conical singularities. We focus on the case of the tetrahedron, for which we have a good
computational control of its geodesics, and compare its curvature profile to that of a smooth sphere. The two are distinct, but
qualitatively similar, which confirms that the curvature profile has averaging properties which are interesting from a quantum point of view.

\vspace{12pt}
\noindent


\newpage

\section{On the nature of quantum observables}
\label{sec:intro}

The difficulty of relating results obtained in nonperturbative, background-inde\-pen\-dent quantum gravity to those found in a
semiclassical regime has many facets. An important one is that the quantum observables currently used to characterize
physics near the Planck scale are different from the typical quantities studied in the context of
perturbative quantum fields on a fixed, curved spacetime, the standard setting for describing the early
universe, say. However, it is not inconceivable that points of contact can be established,
despite the very different nature of the underlying theoretical formalisms. For instance, particular nonperturbative ob\-ser\-vables may 
on sufficiently large scales exhibit semiclassical behaviour that can in principle be captured by quantum field theory on a fixed background.
Setting up such quantitative comparisons could provide valuable mutual con\-sistency checks and
help us identify interesting quantum signatures.  

The concrete context we will use to illustrate the nature of quantum observables is that of 
quantum gravity in terms of Causal Dynamical Triangulations (CDT), where significant progress has been made on the issue. 
CDT quantum gra\-vity is a modern implementation of lattice gravity\footnote{see \cite{livrev} for an overview of earlier
approaches},
based on a nonperturbative, regularized 
path integral over piecewise flat (triangulated) manifolds (see \cite{review1,review2} for reviews). 
Unlike in the standard continuum
formulation of gravity using metrics or vierbeins, these spacetimes can be described in purely geometric terms, 
without introducing
coordinates and their associated gauge redundancy, which is related to the action of the diffeomorphism group.  
This is enormously significant because it \textit{resolves} the vexed problem of 
gauge-fixing the gravitational path integral at
the nonperturbative level. On the other hand, coordinates are gone for good and not easily re-introduced into the quantum
theory, even if one would like to do so for convenience.\footnote{see \cite{toruscoord} for an attempt to introduce coordinates in 
CDT quantum gravity on a four-torus} 
However, even if we did have a global coordinate system valid for all path integral configurations (which we do not),  
the concept of ``a quantity $Q(x)$ at a given point $x$" lacks a physical, label-invariant interpretation due to the
absence of a preferred background metric that could provide a definite frame of reference.  
 
The absence -- in a strongly quantum-fluctuating, Planckian realm -- of some of the ``nice" features of
General Relativity should not be surprising, nor is it specific to the CDT approach. 
It is perfectly compatible with a scenario where even in an extreme quantum regime gravity remains 
essentially geometric in nature, albeit in a way that goes beyond the
differentiable Lorentzian manifolds of the classical theory. CDT quantum gravity is a case in point:
although tensor calculus is not part of its toolbox, it has well-defined notions
of (geodesic) distance and volume, which enter into the construction of geometric quantum observables. 

Generally speaking, it is difficult to find quantum observables that are finite and well-defined in a
nonperturbative regime, and up to now only a handful have been constructed. Most prominent in pure 
quantum gravity without matter coupling are the global Hausdorff dimension \cite{CDT1,CDT2}, 
the spectral dimension \cite{spectral,reconstructing}, the global shape (the so-called 
volume profile) \cite{desitter1,desitter2} and the averaged quantum Ricci curvature \cite{qrc1,qrc2,qrc3}.
In line with our remarks above they are all global quantities, involving either spatial or spacetime integrals.  
Lastly, when comparing different formalisms it should be taken into account that in four dimensions one has little analytical 
control over the path integral.
The expectation values of nonperturbative quantum observables must be
determined with the help of Monte Carlo simulations and one therefore must make sure that
in a given lattice regime they can be measured with sufficient accuracy to
yield reliable results.    

Taken together, the characteristics and construction principles of nonperturbative quantum observables mean
that they are in some sense maximally removed from how we describe the invariant geometric properties of 
classical spacetimes in General Relativity,
namely, in terms of a metric tensor $g_{\mu\nu}(x)$ modulo gauge\footnote{more precisely, in terms of equivalence
classes $[g_{\mu\nu}(x)]$ of Lorentzian metrics under spacetime diffeomorphisms} and local curvature invariants derived from it. 
We will advocate here to partially close this gap by studying a new type of global geometric observable \textit{at the classical
level}. It is a diffeomorphism invariant of the classical theory, taking the form of a spacetime integral over 
a quasi-local geometric quantity\footnote{For simplicity, we assume spacetime to be compact without boundaries, which
covers the standard CDT set-ups with spherical or toroidal spatial slices and a cyclically identified time direction.}, 
which at the same time has a direct analogue in the nonperturbative quantum
theory. Specifically, we will be interested in what we call the \textit{``curvature profile"} of a given spacetime.
This observable is based on the recently introduced \textit{quantum Ricci curvature}, a notion of Ricci curvature
applicable in both classical and Planckian regimes \cite{qrc1,qrc2,qrc3}.

One may wonder how much interesting physical information is conveyed by observables that are spacetime averages,
and of which we have nothing like a complete set, however defined. Such quantities would not be particularly useful
in a classical context where one is interested in resolving the local curvature structure associated with a general matter
distribution. However, our primary motivation is the deep quantum regime, where the universe itself is small (at least the one we 
have access to in computer simulations) and where we have nontrivial indications that the quantum ground state on sufficiently
large scales resembles a four-dimensional de Sitter space, both in terms of its overall shape \cite{desitter1,desitter2} and its averaged 
curvature \cite{qrc3}. If the quantum geometry indeed turns out to be approximated by a de Sitter 
space in a suitably coarse-grained sense, with some accompanying degree of homogeneity and isotropy, 
spacetime-averaged observables may be well suited to investigate its structure on sufficiently large scales. 

As will be reviewed in Sec.\ \ref{sec:qrc} below, the quantum Ricci curvature at a point $x$, which we integrate to obtain the 
curvature profile, is a generalized, quasi-local Ricci curvature associated with both a direction and a geodesic length scale $\delta$. 
The variable $\delta$ 
characterizes the linear size of the neighbourhood around $x$ that contributes to the curvature and can be 
thought of as a coarse-graining scale. It takes values between 0 and the diameter of the manifold. 
After integration, one therefore obtains
a whole function's worth of integrated curvature information, the $\delta$-dependent \textit{curvature profile}.

From the perspective of quantum gravity, classical curvature profiles serve as benchmarks for interpreting the
corresponding quantum curvature profiles (measured in terms of their expectation values), at least for sufficiently large distances $\delta$. 
Since our understanding of curvature profiles of curved classical manifolds is currently limited to smooth constant-curvature
spaces\footnote{of Riemannian signature, to match the Wick-rotated results of CDT quantum gravity} and
Delaunay triangulations of such spaces \cite{qrc1}, 
an important step is to get
a better understanding of how these profiles encode the geometric and topological properties of a larger variety of classical
spaces. A natural class of spaces to consider are those where the maximal isometry of a constantly curved space 
is partially broken, leaving a space that is still sufficiently simple to allow for control of its geodesics, which is needed
for the computation of the quantum Ricci curvature. 

The work presented here examines a class of two-dimensional, compact model spaces of non-negative curvature, where 
the rotational $SO(3)$-symmetry of the constantly curved sphere is broken to a residual discrete subgroup associated
with the symmetries of a regular, convex polyhedron. A primary aim is to quantify how the \textit{distribution of curvature} is
reflected in the curvature profiles of the corresponding spaces, with the two-dimensional case chosen for simplicity. 
According to the Gauss-Bonnet theorem, the total
Gaussian curvature of any two-dimensional space of spherical topology is given by $4\pi$. On a round two-sphere,
this curvature is distributed completely evenly. The surface of a regular tetrahedron represents the
opposite extreme: it is flat almost everywhere, except for four isolated, singular points, each associated with a Gaussian
curvature (deficit angle) of $\pi$. There are two main questions which we will try to answer in what follows:
can we tell the two spaces apart by comparing their volume profiles? And how
effective is the averaging in ``smearing out" the effect of the isolated singularities?

In the following Sec.\ \ref{sec:qrc}, we recall some key formulas for computing the quantum Ricci curvature on
continuum geometries in terms of sphere distances and define the notion of a curvature profile. 
In Sec.\ \ref{sec:cone}, we analyze the influence of an isolated conical singularity on the average sphere distance,
which is a crucial ingredient in the curvature measurements. It allows us to compute the curvature profiles 
at short distance scales $\delta$ of a class of regular polyhedral surfaces associated with Platonic solids in Sec.\ \ref{sec:measure}.
In Sec.\ \ref{sec:tetra} we discuss the construction of geodesics
on the surface of a regular tetrahedron, which is needed to determine the geodesic circles appearing in the
curvature construction. This turns out to be a nontrivial task, which can be tackled by unfolding the surface onto the
flat plane. It enables us to determine geo\-desic circles of arbitrary location and radius, which then serves as an input
for the numerical computation of the curvature profile. 
The final Sec.\ \ref{sec:final} contains a summary and our conclusions.

\section{Quantum Ricci curvature and curvature profiles}
\label{sec:qrc}

The motivation for introducing the quantum Ricci curvature \cite{qrc1} was the need for a well-defined notion of renormalized curvature 
in nonperturbative quantum gravity and CDT quantum gravity in particular. It was inspired by a generalized notion of Ricci
curvature due to Ollivier \cite{ollivier} and has also been used in graph-theoretic models of quantum gravity
\cite{tru1,tru2,tru3,tru4}. Its classical starting point is the observation that two sufficiently close and sufficiently small geodesic spheres 
on a positively
curved Riemannian space are closer to each other than their respective centres, 
while the opposite is true on a negatively curved space. This leads to the idea of \textit{defining} curvature on more general
(e.g.\ nonsmooth) metric spaces by comparing the distances of spheres with the distances of their centres.

The \textit{quantum Ricci curvature} is a particular implementation of this idea, designed specifically for 
use on the ensembles of piecewise flat simplicial configurations of nonperturbative, dynamically triangulated
quantum gravity models.
It has been applied to Euclidean dynamical triangulations in two dimensions \cite{qrc2} and more recently
to the physically relevant case of CDT in four dimensions \cite{qrc3}, demonstrating its viability in quantum gravity. 
However, the prescription can in a straightforward
way be implemented in other geometric settings, as long as one has notions of distance and volume. This includes classical continuum 
spaces, which we are focusing on presently as part of an effort to build up a reference catalogue of
curvature profiles, for comparison with quantum results.

\begin{figure}[h]
\centerline{\scalebox{0.45}{\rotatebox{0}{\includegraphics{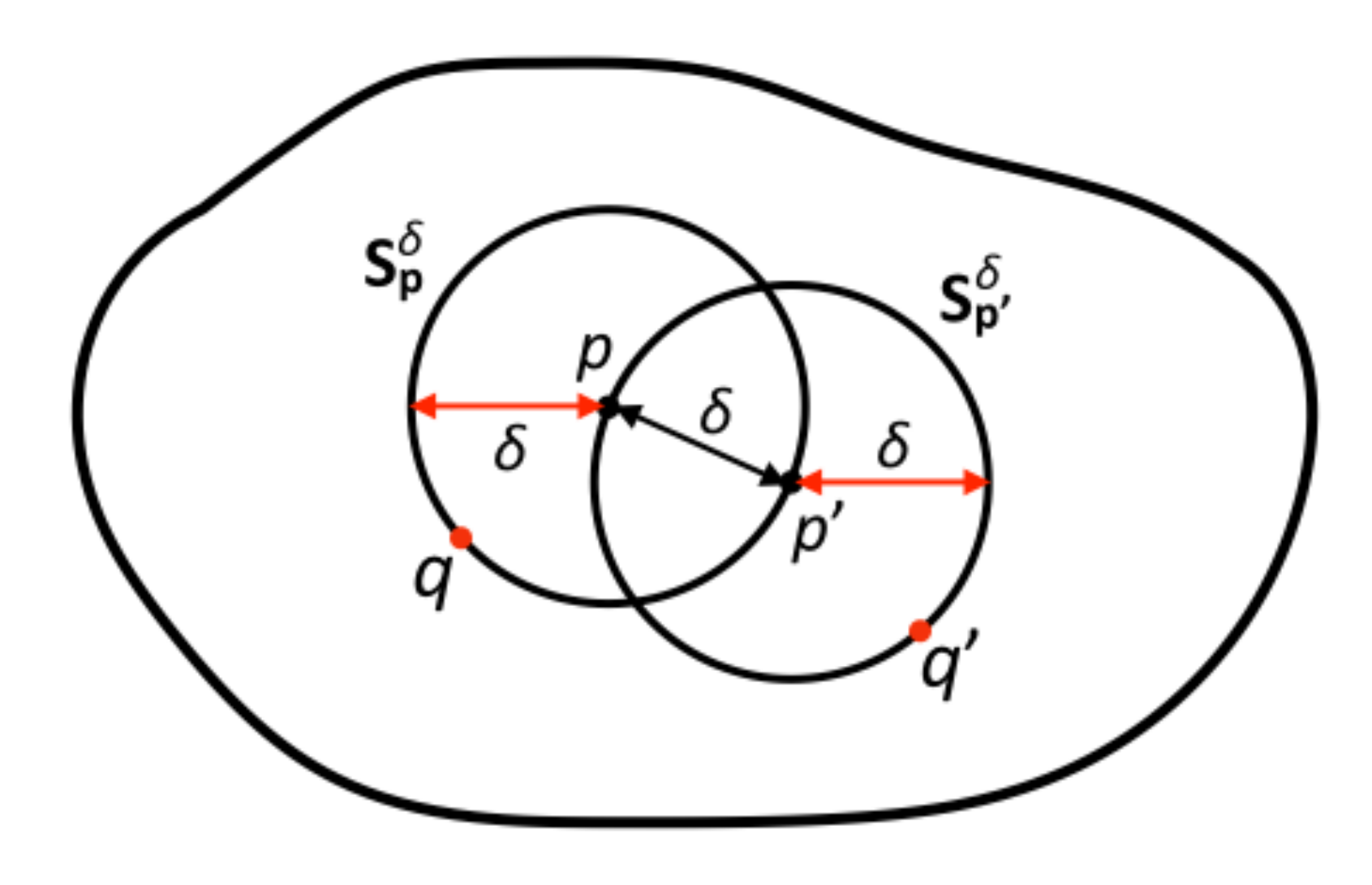}}}}
\caption{The quantum Ricci curvature $K(p,p')$ is extracted from the computation of the average sphere distance (\ref{sdist})
of two overlapping spheres $S_p^\delta$ and $S_{p'}^\delta$ of radius $\delta$, whose centres $p$ and $p'$ are a distance $\delta$ apart.}
\label{fig:newintersect}
\end{figure}

Specializing to a two-dimensional continuum space $M$ with metric $g_{\mu\nu}(x)$ and associated geodesic distance $d_g$,
the quasi-local set-up associated with 
the quantum Ricci curvature $K(p,p')$ consists of two intersecting geodesic circles (``one-spheres")
$S_p^\delta$ and $S_{p'}^\delta$ of radius $\delta$, with centres $p$ and $p'$ a distance $\delta$ apart
(Fig.\ \ref{fig:newintersect}).
To extract $K(p,p')$ one computes the \textit{average sphere distance}\footnote{one could also call it the average \textit{circle} 
distance, but we will keep using the more general term in what follows} $\bar{d}(S_p^{\delta},S_{p'}^{\delta})$,  
which by definition is given by the normalized double-integral 
\begin{equation}
\bar{d}(S_p^{\delta},S_{p'}^{\delta}):=\frac{1}{\textit{vol}\,(S_p^{\delta})}\frac{1}{\textit{vol}\,(S_{p'}^{\delta})}
\int_{S_p^{\delta}}dq\; \sqrt{h} \int_{S_{p'}^{\delta}}dq'\; \sqrt{h'}\ d_g(q,q')
\label{sdist}
\end{equation}
over all pairs of points $(q,q')\in S_p^\delta \times S_{p'}^\delta$,
that is, by \textit{averaging} the distance $d_g(q,q')$ over the two circles,
where $h$ and $h'$ are the determinants of the induced metrics on $S_p^{\delta}$ and $S_{p'}^{\delta}$,
which are also used to compute the one-dimensional circle volumes $vol(S)$.
From eq.\ (\ref{sdist}), the quantum Ricci curvature $K(p,p')$ associated with the point pair $(p,p')$ 
is defined by
\begin{equation}
\bar{d}(S_p^{\delta},S_{p'}^{\delta})/\delta=c\, (1 - K(p,p')),\;\;\;\;\;\;\;\; \delta =d_g(p,p'),
\label{qric}
\end{equation}
where the prefactor $c$ is given by $c\! :=\!\lim_{\delta\rightarrow 0} \bar{d}/\delta$, and in general may depend
on the point $p$.
On smooth Riemannian manifolds and for small distances $\delta$, one can expand the quotient (\ref{qric}) 
into a power series in $\delta$, with $c$ a constant that only depends on the dimension. In two dimensions,
one finds \cite{qrc2}
\begin{equation}
\bar{d}/\delta= 1.5746+\delta^2 \left(-0.1440\, \textit{Ric}(v,v)+{\cal O}(\delta)\right),
\label{2dexp}
\end{equation}
where $\mathit{Ric}(v,v)\! =\! R_{ij}v^iv^j$ is the usual Ricci curvature, evaluated on the unit vector $v$ at the point $p$ 
in the direction of $p'$, and the numerical coefficients are rounded to the digits shown. 

To obtain a global observable of the type discussed in Sec.\ \ref{sec:intro}, we integrate the average sphere
distance (\ref{sdist}) over all positions $p$ and $p'$ of the two circle centres, 
while keeping their distance $\delta$ fixed. The spatial average over the average sphere distance at
the scale $\delta$ is 
\begin{equation}
\bar{d}_{\rm av}  (\delta):=\frac{1}{Z_\delta}
\int_{M}d^2x \sqrt{g} \int_M d^2 x' \sqrt{g}\;\,  \bar{d}(S_x^{\delta},S_{x'}^{\delta})\; \delta_D(d_g(x,x'),\delta),
\label{avdist}
\end{equation}   
where $\delta_D$ denotes the Dirac delta function and the normalization factor $Z_\delta$ is given by
\begin{equation}
Z_\delta=\int_{M}d^2x \sqrt{g} \int_M d^2 x' \sqrt{g}\;\; \delta_D(d_g(x,x'),\delta).
\label{zdelta}
\end{equation}
The integration in eq.\ (\ref{avdist}) includes an averaging over directions, which means that it will
allow us to extract an (averaged) \textit{quantum Ricci scalar} $K_{\rm av}(\delta)$. 
The curvature profile is now given by the quotient
\begin{equation}
\bar{d}_{\rm av}(\delta)/\delta=:c_{\rm av} (1 - K_{\rm av}(\delta)),
\label{profile}
\end{equation}
where the constant $c_{\rm av}$ is defined by $c_{\rm av}\! :=\!\lim_{\delta\rightarrow 0} \bar{d}_{\rm av}/\delta$.
Note that for the special case where $M$ is the round two-sphere no spatial averaging is necessary to obtain the curvature
profile \cite{qrc1}, because in that case the average sphere distance (\ref{sdist})
depends only on the distance $\delta$, and not on the locations of $p$ and $p'$. 

Although in the above discussion we have concentrated on the case of two dimensions, an analogous construction goes through in 
higher dimensions too. As emphasized in the introduction, the classical curvature profile (\ref{profile}) 
can be translated directly into a quantum observable,
whose expectation value $\langle \bar{d}_{\rm av}(\delta)/\delta \rangle$ can 
be determined numerically in an approach like CDT quantum gravity. 
In such a lattice approach, all distances are given in dimensionless
lattice units, which can be converted into dimensionful units invoking the lattice spacing $a$, an ultraviolet length cutoff
that is taken to zero in any continuum limit. 
From the dimensionless curvature $K_{\rm av}(\delta)$ measured in the nonperturbative quantum theory
one can extract a dimensionful, renormalized quantum Ricci scalar 
$K^r\! (\delta_{\rm ph})$, which depends on a physical coarse-graining or renormalization scale $\delta_{\rm ph}\! :=\! a\delta$, 
via
\begin{equation}
K_{\rm av}(\delta)=:\delta^2 a^2 K^r\! (\delta_{\rm ph})=(\delta_{\rm ph})^2 K^r\! (\delta_{\rm ph}), 
\label{qren}
\end{equation}
in the limit $a\rightarrow 0$. However, as discussed in the introductory section, in this work we will only be interested in the curvature
profiles of certain classical spaces, where lengths and volumes vary continuously without any short-distance cutoff.
As a preparation, we will in the next section examine the influence of a conical singularity on the average sphere
distance (\ref{sdist}) in two dimensions.

\section{The influence of conical singularities}
\label{sec:cone}

From the point of view of its intrinsic geometric properties, the surface of a regular tetrahedron is flat everywhere, apart from its
corners, where curvature is concentrated in the manner of a delta function. The curvature that appears along its edges (with the
exception of the four corners) in a
standard embedding picture of the tetrahedron in Euclidean $\R^3$ is only extrinsic and not of interest in our present context.
The same is true for the surfaces of other regular polyhedra. It implies that the intrinsic geometry of the 
neighbourhood of any of their corners is identical to that of a cone with a conical curvature singularity at its ``tip". 
To obtain the curvature profile of such a classical
two-dimensional space, we must understand how the presence of one or more conical singularities affects the average sphere
distance of a pair of circles on them. 

Our quantitative assessment starts by examining the influence of a single conical
singularity. Clearly, since the metric of a cone is flat everywhere except at the point where the singularity is located,
if the pair $(S_p^{\delta},S_{p'}^{\delta})$ of $\delta$-circles depicted in Fig.\ \ref{fig:newintersect} is sufficiently far away
from the singularity, the normalized average sphere distance is that of flat space, 
$\bar{d}(S_p^{\delta},S_{p'}^{\delta})/\delta\!\approx \! 1.5746$. On the other hand, if the conical singularity lies
somewhere inside the double circle, one would expect it to have a nontrivial effect. However, somewhat contrary to
na\"ive expectation, the curvature singularity has an influence on the average sphere distance even when it lies
strictly outside the two circles, as long as it is sufficiently close to them. As we will show in more detail below, this is due to the 
presence of ``geodesic shortcuts" associated with the conical singularity.

\begin{figure}
\centering
\begin{subfigure}[t]{0.45\textwidth}
\centering
\includegraphics[height=0.6\linewidth]{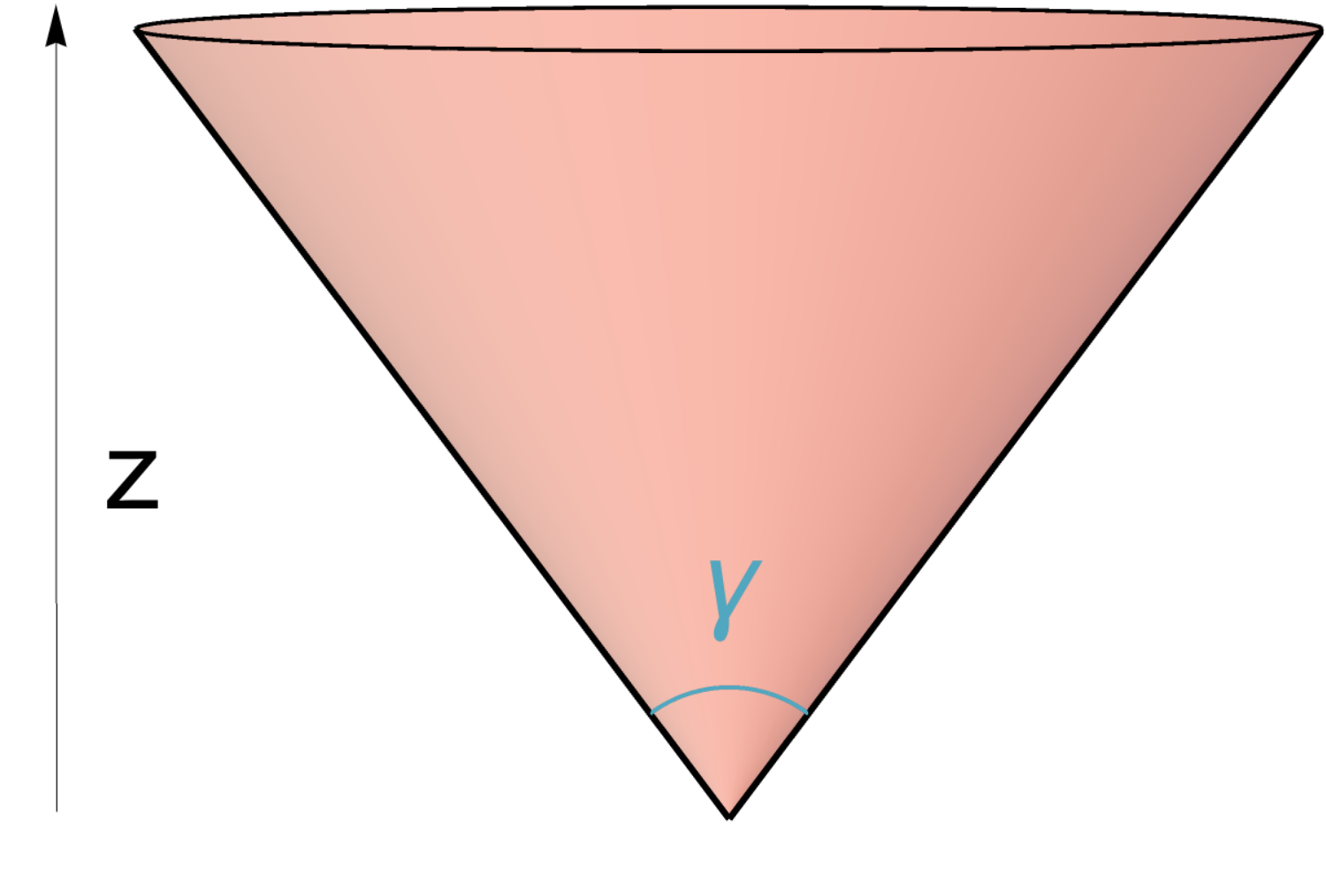}
\end{subfigure}
\hspace{0.05\textwidth}
\begin{subfigure}[t]{0.45\textwidth}
\centering
\includegraphics[height=0.6\linewidth]{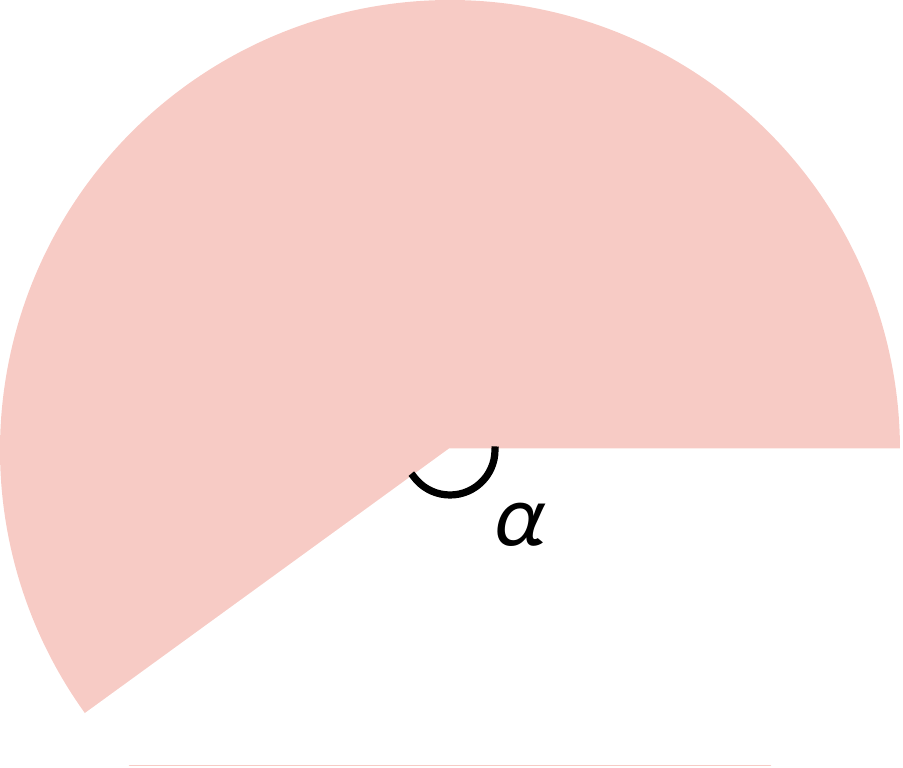}
\end{subfigure}
\caption{A geometry with a conical singularity can be represented as a
cone in Euclidean space with apex angle $\gamma$  (left) 
or a flat plane with an angle $\alpha$ removed (right).}
\label{fig:cones}
\end{figure}

\subsection{Geometry of the cone}

One way of describing a cone with apex angle $\gamma$ is as the set of all points $(x,y,z)\in\R^3$ satisfying 
$z\! =\!\cot (\gamma/2) \sqrt{x^2+y^2}$, where $0\! <\!\gamma\!\leq\!\pi$ (Fig.\ \ref{fig:cones}, left).
The apex or tip of the cone lies at the origin of the Cartesian
coordinate system, and $\gamma$ is the angle formed in the plane $x\! =\! 0$ by its intersection with the cone.
A point on the cone can be labelled by its Euclidean distance $r$ to the origin and a rotation angle 
$\theta =\arctan y/x \in [0,2\pi]$. The induced metric on the cone in terms of these coordinates is given by
\begin{equation}
ds^2 = dr^2 + \sin^2 ( \tfrac{\gamma}{2})\, r^2\,d\theta^2.
\label{eq:metric}
\end{equation}
An alternative description of the cone is obtained by cutting out an infinite wedge with angle $\alpha$ 
from the Euclidean plane $\R^2$ (Fig.\ \ref{fig:cones}, right) and identifying boundary points pairwise across the wedge. 
The result is a cone with a curvature singularity characterized by the deficit angle $\alpha$, which is
a direct measure of Gaussian curvature. Since we are only interested in singularities with positive 
curvature, the relevant angle range is $0\! \leq \! \alpha\!<\! 2\pi$. The relation with the apex angle $\gamma$
is $\alpha = 2\pi (1-\sin (\gamma/2))$. 

The useful feature of the ``planar" representation of the cone in terms of a part of 
$\R^2$ is the fact that geodesics are simply given by straight lines. A knowledge of geodesics is
needed to determine the geodesic distances between points.\footnote{Here and in the remainder of the paper,
a geodesic is defined as a locally shortest curve that does not contain any singularities, with the possible exception of
its endpoints.}
The only minor difficulty one has to take care of
is to continue a geodesic correctly across the wedge if necessary. Placing the singularity at the origin of $\R^2$ and
using standard spherical coordinates $(r,\varphi)$, the metric on the cone in this parametrization is
\begin{equation}
ds^2=dr^2+r^2 d\varphi^2,
\label{eq:flat}
\end{equation}
where the range of the angle $\varphi$ is limited to $\varphi\in [0,2\pi-\alpha]$. Using the notation $r$ is justified, because
the radial geodesic distance here is identical to the one we used for the cone embedded in $\R^3$. Comparing the
two metrics (\ref{eq:metric}) and (\ref{eq:flat}), we see that they are simply related by a constant rescaling of the 
angles $\theta$ and $\varphi$, namely, $\varphi = \sin(\gamma/2) \theta$. 
The distance between two points $(r_1,\varphi_1)$ and $(r_2,\varphi_2)$ is simply their Euclidean distance
\begin{equation}
d\left( (r_1,\varphi_1), (r_2,\varphi_2) \right) = \sqrt{r_1^2+r_2^2-2 r_1 r_2 \cos (\varphi_2 -\varphi_1) },
\label{eudist}
\end{equation}
whenever the absolute angle difference $|\varphi_2-\varphi_1|$ does not exceed $\pi -\alpha/2$. If it does,
the argument of the cosine should be substituted by the angle $2\pi-\alpha-|\varphi_2 -\varphi_1|$. 
By rescaling the angles, we can immediately derive a distance function for points on the cone labelled by $r$ and $\theta$, namely,
\begin{equation}
 d\left( (r_1,\theta_1), (r_2,\theta_2) \right) = \sqrt{r_1^2+r_2^2-2 r_1 r_2 \cos (\sin(\tfrac{\gamma}{2}) \Delta (\theta_1,\theta_2)) },
\label{conedist}
\end{equation} 
where
\begin{equation}
\Delta (\theta_1,\theta_2):= \min (|\theta_2-\theta_1|, 2\pi- |\theta_2-\theta_1| ).
\label{Delta}
\end{equation}
In what follows, we will work with the cone parametrization in terms of coordinates $(r,\theta)$, which is more
convenient in the applications we will consider.

\begin{figure}
\centering
\begin{subfigure}[t]{0.47\textwidth}
\centering
\includegraphics[height=0.9\linewidth]{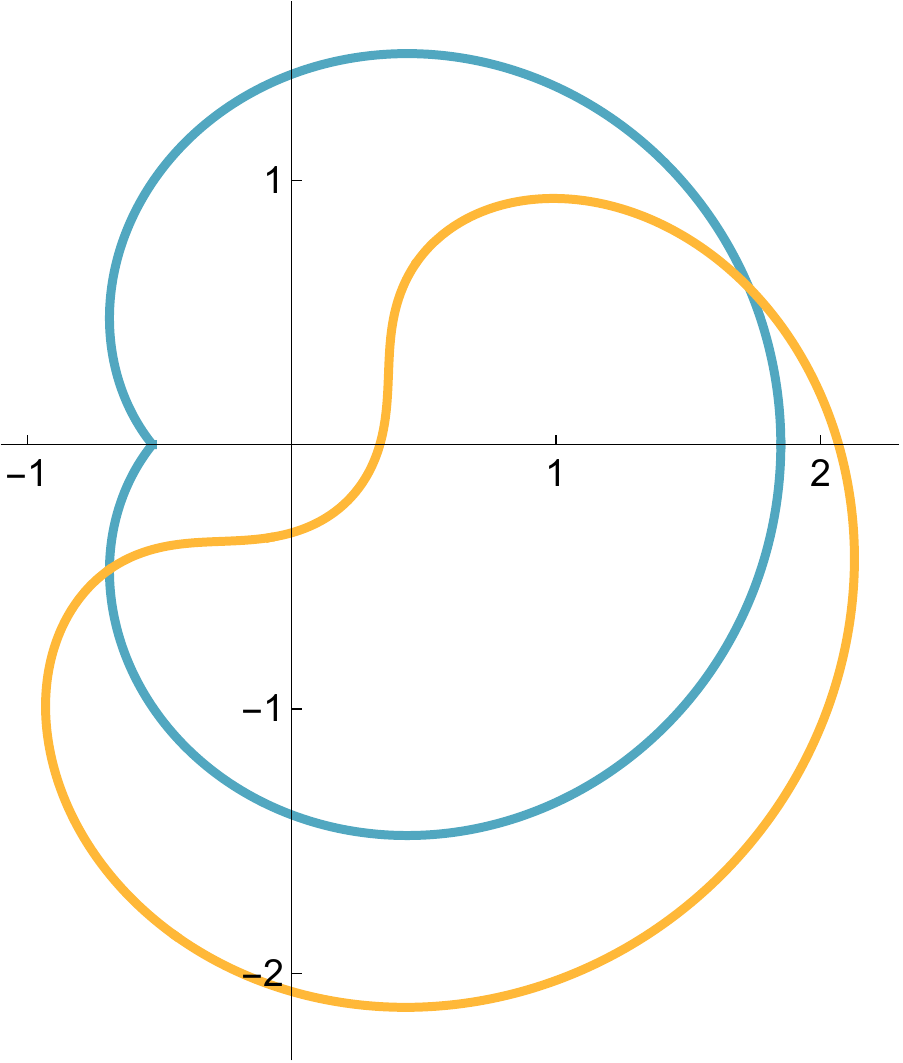}
\end{subfigure}
\hspace{0.04\textwidth}
\begin{subfigure}[t]{0.47\textwidth}
\centering
\includegraphics[height=0.9\linewidth]{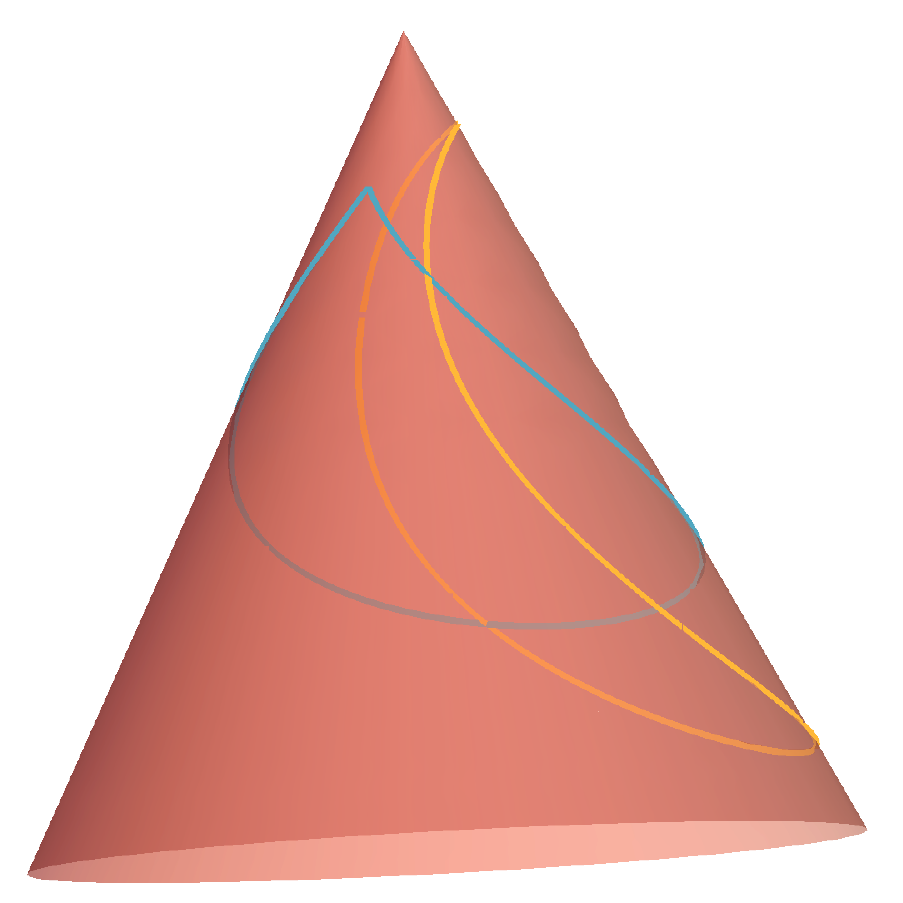}
\end{subfigure}
\caption{Two unit circles in the $(r, \theta)$-plane with a singularity at the origin (left), 
and on the cone embedded in $\mathbb{R}^3$ (right). The blue circle encloses the singularity and shows a discontinuous jump of its tangent vector at $\theta = \pi$.}
\label{fig:tangent-vector-jump}
\end{figure}

\subsection{Computing average sphere distances}
\label{subsec:sd}

To compute the average sphere distances (\ref{sdist}) on the cone $M$, we must first parametrize 
geodesic circles on $M$, where the geodesic circle of radius $\delta$ centred at the point $p$ consists of all points $q$ at distance 
$\delta$ from $p$, $S_p^\delta =\{q\in M |\, d(p,q)\! =\! \delta\}$. 
Let us first discuss the parametrization of a geodesic
circle which encloses the singularity at the origin. This case is slightly simpler, because the angle $\theta$ of $M$
can be used to label the points of $S_p^\delta$ uniquely. Using the rotational invariance of the set-up, we can without
loss of generality choose the centre $p$ to have coordinates $p=(r_1,\theta_1\! =\! 0)$, where $r_1 <\delta$. 
Labelling a point $q$ on the circle
$S_p^\delta$ by $q=(r_2,\theta_2)$, it must by assumption satisfy $d(p,q)\! =\! \delta$, which is a quadratic equation
for $r_2$. Because of relation (\ref{Delta}), we should distinguish between the cases $0\leq\theta_2\leq\pi$ and
$\pi\leq\theta_2\leq 2\pi$. The unique solutions $r_2(\theta_2)$ are
\begin{equation}
r_2 (\theta_2)\!  = \! \begin{cases}
r_1\cos\!\left( \theta_2 \sin (\frac{\gamma}{2})\right)\! +\! \sqrt{\delta^2\! -\! r_1^2 
\sin^2\!\left( \theta_2\sin (\frac{\gamma}{2})\right)}, \;\;\; & 0 \leq\theta_2\leq \pi , \\
r_1\cos\!\left( (2\pi\! -\!\theta_2) \sin (\frac{\gamma}{2})\right)\! +\! \sqrt{\delta^2\! -\! r_1^2 \sin^2\!
\left( (2\pi\! -\!\theta_2)\sin (\frac{\gamma}{2})\right)},
 & \pi\leq\theta_2\leq 2\pi .
\end{cases}
\label{r2second}
\end{equation}
They can be used to uniquely parametrize the points along $S_p^\delta$ by the curve
\begin{equation}
c^\mu(\theta)=(r_2(\theta),\theta),\;\;\;\;\; 0\leq \theta \leq 2\pi ,
\label{curve1}
\end{equation}
with curve parameter $\theta$, and where for ease of notation we have suppressed the dependence of the curve on $p$ and $\delta$.
Note that this curve is smooth everywhere apart from the point $\theta\! =\!\pi$,
where its tangent vector $dc^\mu/d\theta$ jumps in a discontinuous way, resulting in a kink in the 
curve $c(\theta)$ (Fig.\ \ref{fig:tangent-vector-jump}).
If the centre $p$ has coordinates $(r_1,\theta_1)$, for some nonvanishing angle $\theta_1$, the angle
$\theta_2$ on the right-hand sides of the relations (\ref{r2second}) should be substituted by $| \theta_2 -\theta_1|$.

The parametrization of a geodesic circle which does not enclose the curvature singularity can also be parametrized by
a $\theta$-angle, but not uniquely. As we will see below, a convenient choice in any integration along $S_p^\delta$ 
is to parametrize the circle along two separate segments and then add their contributions. 
Again we choose the point $p=(r_1,0)$ as the centre of the circle, whose radial
coordinate now satisfies $r_1 \geq\delta$. When solving the equation $d(p,q)\! =\! \delta$ for $r_2$ as before, there are
two differences. First, the angle $\theta_2$ can only
vary in the interval $[-\theta_{\rm m},\theta_{\rm m}]$, where the maximal angle $\theta_m$ is defined by
$\sin (\sin{ (\gamma/2)}\, \theta_{\rm m})\! =\! \delta/r_1$.
This is easy to see by examining the geometry of the situation in the flat-space coordinates $(r,\varphi)$.  
Second, there are two solutions $r_2(\theta_2)$ for every value $\theta_2$ in the interior of this interval, namely,
\begin{equation}
r^\pm_2(\theta_2)=
r_1\cos\!\left( \theta_2 \sin (\tfrac{\gamma}{2})\right)\! \pm \! \sqrt{\delta^2\! -\! r_1^2 
\sin^2\!\left( \theta_2\sin (\tfrac{\gamma}{2})\right)},
\;\; -\theta_{\rm m} \leq\theta_2\leq \theta_{\rm m}. 
\label{r2both}
\end{equation}
One easily verifies that the argument of the square root cannot become negative in the angle range considered.
A straightforward way to parametrize the points $q$ along the circle is by splitting the circle into two segments 
$c_+(\theta)$ and $c_-(\theta)$,
corresponding to the two solutions (\ref{r2both}), namely,
\begin{eqnarray}
&c_+^\mu(\theta) = \left(r_2^+(\theta), \theta \right), \;\;&  -\theta_{\rm m} \leq \theta \leq  \theta_{\rm m}, \\
&c_-^\mu(\theta) = \left(r_2^-(\theta), \theta \right), \;\; &  -\theta_{\rm m} \leq \theta \leq \theta_{\rm m}.
\label{eq:circle-param-nenc}
\end{eqnarray}
If the angle $\theta_1$ of the point $p\! =\! (r_1,\theta_1)$ is nonvanishing, the angles $\theta_2$ in the above
expressions should again be substituted appropriately.
As a cross check of the circle parametrizations, one can use them to compute the lengths (one-dimensional volumes) of the circles,
which at any rate are needed in the computation of the average sphere distances (\ref{sdist}). For the case that the circle encloses
the singularity, one finds
\begin{align}
\hspace{-1.2cm} vol (S^\delta_p) & =\int_0^{2 \pi}\!\!\! d\theta\, \sqrt{g_{\mu\nu} \tfrac{dc^\mu}{d\theta} \tfrac{dc^\nu}{d\theta}} \nonumber \\
&= \int_0^\pi \! d\theta\, \delta \sin(\tfrac{\gamma}{2}) (1\! +\! {\cal R}(\theta))+
\!\int_\pi^{2 \pi} \!\!\! d\theta\, \delta \sin(\tfrac{\gamma}{2}) (1\! +\! {\cal R}(2\pi -\theta))\\
&=2\pi\delta \sin (\tfrac{\gamma}{2})+2\delta \arcsin \left( \tfrac{r_1}{\delta}\sin \left( \pi \sin (\tfrac{\gamma}{2} )  \right)  \right),
\hspace{2.5cm} [r_1 <\delta ] \nonumber 
\label{vol1}
\end{align}
where $g_{\mu\nu}$ refers to the metric (\ref{eq:metric}) and we have introduced the shorthand notation
\begin{equation}
{\cal R}(\theta)= \frac{r_1 \cos\left( \theta \sin (\tfrac{\gamma}{2}) \right)}{\sqrt{\delta^2-r_1^2 \sin^2 \left( \theta \sin (\tfrac{\gamma}{2}) \right)}}.
\label{rshort}
\end{equation}
For the case that the circle does not enclose the singularity, the corresponding computation yields
\begin{align}
\hspace{-1.1cm} vol (S^\delta_p) & =\int_{-\theta_{\rm m}}^{\theta_{\rm m}}\!\!\! d\theta\, 
\sqrt{g_{\mu\nu} \tfrac{dc^\mu_+}{d\theta} \tfrac{dc^\nu_+}{d\theta}} 
+ \int_{-\theta_{\rm m}}^{\theta_{\rm m}}\!\!\! d\theta\, 
\sqrt{g_{\mu\nu} \tfrac{dc^\mu_-}{d\theta} \tfrac{dc^\nu_-}{d\theta}} \nonumber \\
& =\int_{-\theta_{\rm m}}^{\theta_{\rm m}}\! d\theta\, \delta \sin(\tfrac{\gamma}{2}) (1\! +\! {\cal R}(\theta))+
\!\int_{-\theta_{\rm m}}^{\theta_{\rm m}} \!\!\! d\theta\, \delta \sin(\tfrac{\gamma}{2}) (-1\! +\! {\cal R}(\theta))\\
&=2\pi\delta, \hspace{9cm} [r_1 \geq \delta ] \nonumber 
\label{vol2}
\end{align}
which is the expected flat-space result.\footnote{Note that 
the tangent vectors $dc^\mu_\pm /d\theta$ diverge at the endpoints $\pm \theta_{\rm m}$ of the two segments.
To avoid problems in the numerical implementation of the sphere
distance calculations we removed a tiny interval of width $\sim\ 10^{-8}$ from the integration range around such
endpoints. This does not affect results at the accuracy considered.}

\begin{figure}
\centering{}
\begin{subfigure}[t]{0.49\textwidth}
\hspace{0.05\textwidth}\includegraphics[height=0.55\linewidth]{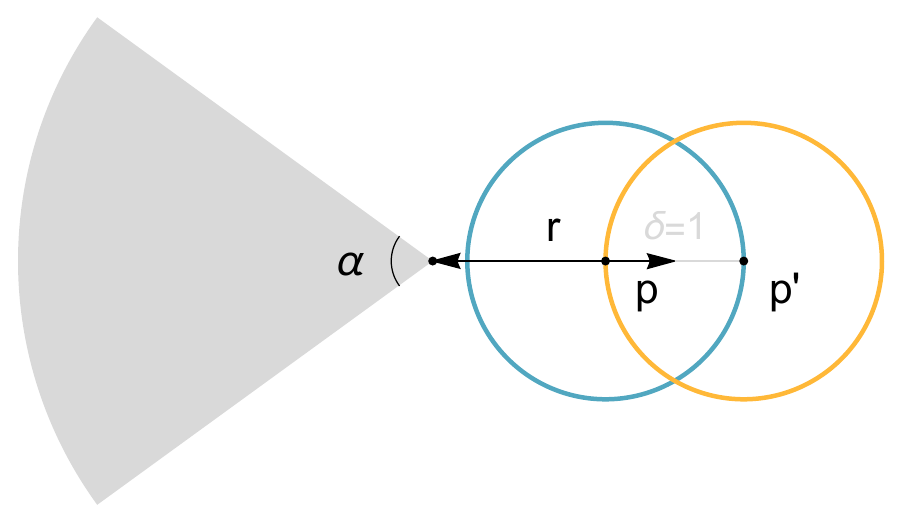} \\ \\
\includegraphics[width=1\linewidth]{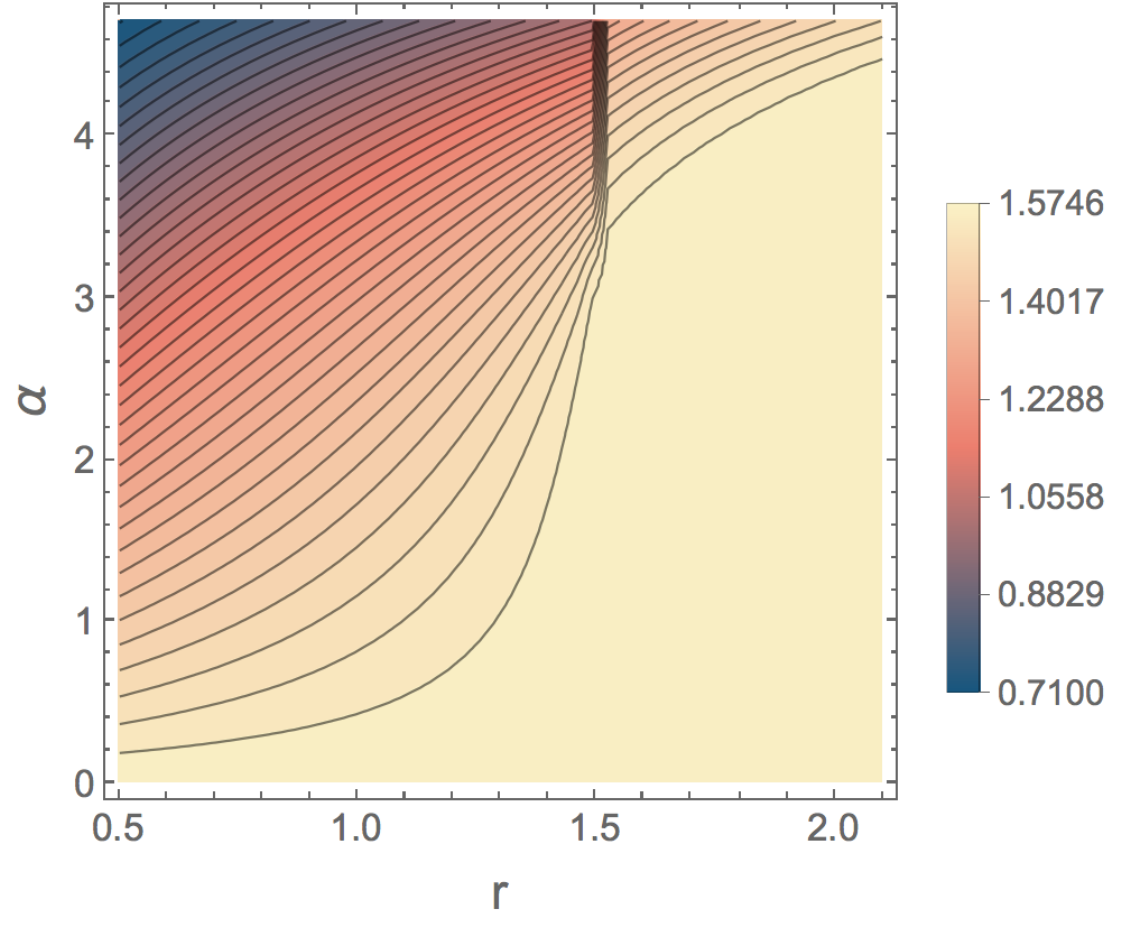}
\end{subfigure}
\begin{subfigure}[t]{0.49\textwidth}
\hspace{0.05\textwidth}\includegraphics[height=0.55\linewidth]{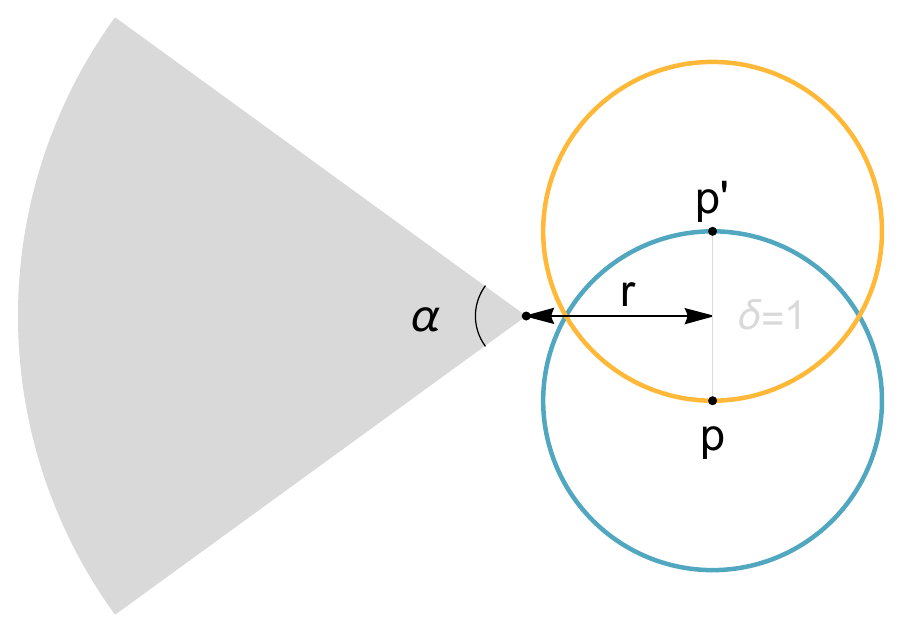} \\ \\
\includegraphics[width=1\linewidth]{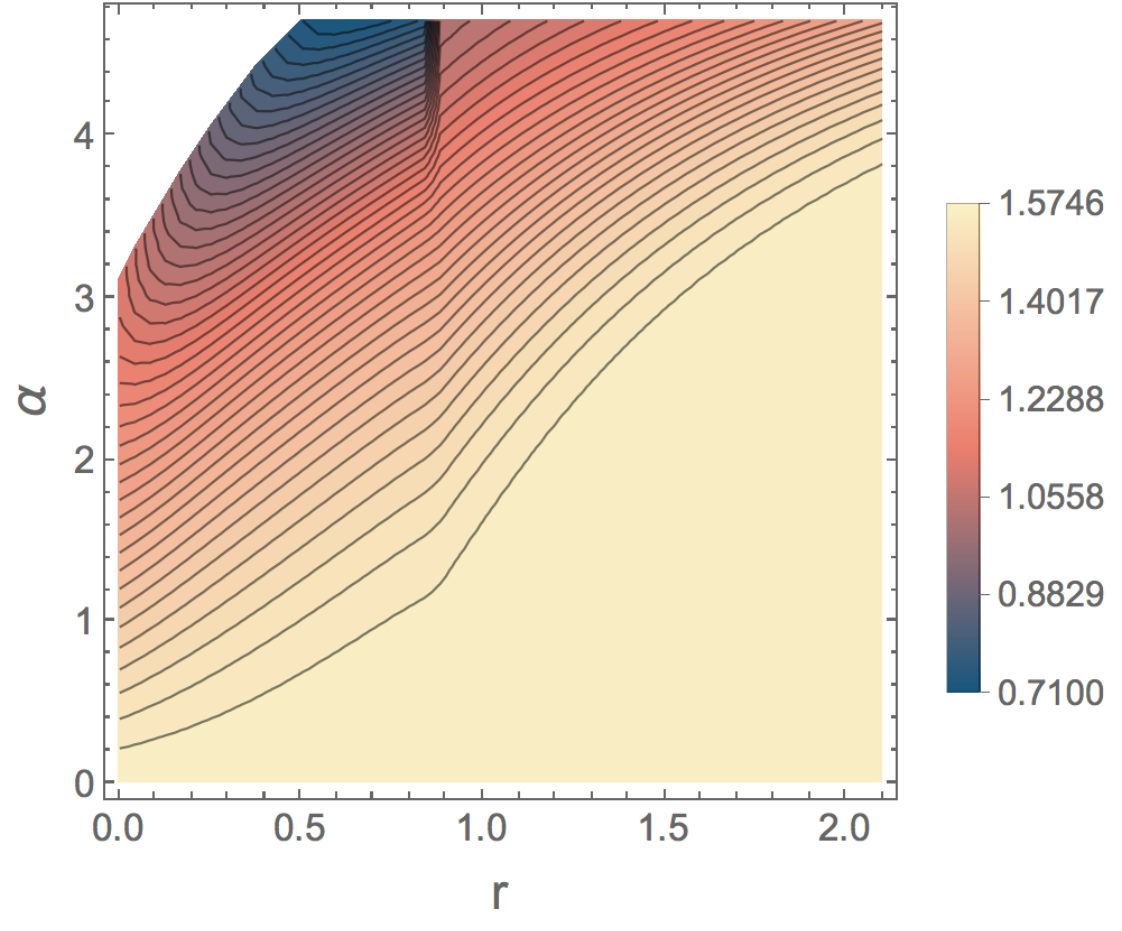}
\end{subfigure}
\caption{Contour plots of the average sphere distance $\bar{d} (S_{p}^\delta,S_{p'}^\delta)$ as a function of $\alpha$ and $r$, 
for two particular circle arrangements,
as described in the text. The distance $r$ is given in units of $\delta$ and the deficit angle $\alpha$ in radians.}
\label{fig:contour-plots}
\end{figure}
We now have all ingredients in hand to explicitly compute the average sphere distance (\ref{sdist}) for 
pairs $(p,p')$ of points at distance $\delta$. 
We perform the double-integral numerically in {\sc Mathematica}, where the intermediate accuracy is kept at 8 significant digits. 
Apart from the strength of the singularity (captured by the deficit angle $\alpha$)
and the linear distance $\delta$, $\bar{d}(S_p^{\delta},S_{p'}^{\delta})$ depends on the distance
of $p$ from the singularity and on the orientation of the second circle relative to the first. 

To illustrate the behaviour of $\bar{d}$,
Fig.\ \ref{fig:contour-plots} shows contour plots for two particular orientations of the double circle relative to the singularity,
where the average sphere distance is given as a function of the deficit angle $\alpha$ and the distance $r$ of the midpoint between
$p$ and $p'$ to the singularity. To fix the overall scale, we have set $\delta\! =\! 1$. Recall that 
for $\alpha\! =\! 0$, there is no singularity, and that in flat space $\bar{d}\!\approx\! 1.5746$. 
As indicated by the darker shades in Fig.\ \ref{fig:contour-plots}, the value of $\bar{d}$
decreases with increasing $\alpha$ and decreasing distance to the singularity, characteristic of a positive and growing
quantum Ricci curvature. The illustrations of the circle positions use the planar representation of the
cone with a wedge removed. In Fig.\ \ref{fig:contour-plots}, left, the circle centres are chosen collinear with the singularity. For $r\! \geq\! 1.5$,
neither of the circles encloses the singularity and there is no or little deviation from the flat-space behaviour, unless $\alpha$
increases beyond $\pi$. In this case the wedge will ``cut into" the first circle, giving rise to a nontrivial effect. The region
$r\! < \! 0.5$ is not well defined, since it would imply that $p$ lies inside the wedge. In Fig.\ \ref{fig:contour-plots}, right, the two circles are
arranged symmetrically with respect to the singularity. What is noteworthy here is the fact that even when the singularity
lies outside the circle configuration, i.e. $r\! >\! \sqrt{3}/2\approx 0.866$, and the size of the deficit angle is only moderate, there is a nontrivial
effect on $\bar{d}$. We will look closer at this phenomenon in the next subsection. The top left-hand corner of the contour
plot is not defined, since both $p$ and $p'$ would lie inside the wedge.

\subsection{Domain of influence}

So far, the developments of this section have been concerned with the geometry in the vicinity 
of an isolated curvature singularity on an infinitely extended cone.
However, we are interested in determining the curvature profiles of compact, flat spaces with
several such singularities, more precisely, the regular polyhedral surfaces of the so-called Platonic solids: the tetrahedron, octahedron,
cube, dodecahedron and icosahedron. The method developed for computing average sphere distances on a cone can be
used on these polyhedra too, but only if the double circles are sufficiently small to not be influenced by more than one of the
conical singularities at the corners of these polyhedra. For a surface with a given area, determined by the length of an edge,
this imposes an upper bound on the size $\delta$ of the circles. Since we need to average over all double circles of size $\delta$ to
obtain the curvature profile, we must determine the size $\delta_{\rm max}$ below which the computation of
$\bar{d} (S_{p}^\delta,S_{p'}^\delta)$ will never be influenced by more than one of the corner points.

\begin{figure}
\centering
\begin{subfigure}[t]{0.45\textwidth}
\includegraphics[width=\linewidth]{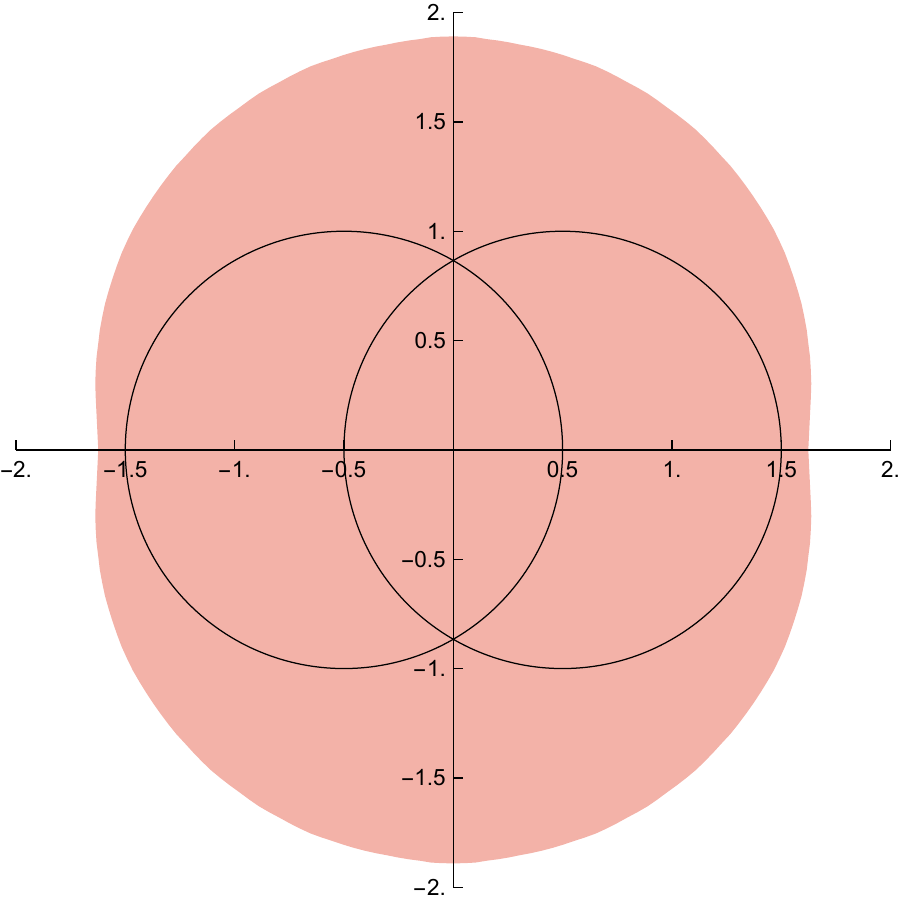}
\end{subfigure}
\hspace{0.02\textwidth}
\begin{subfigure}[t]{0.45\textwidth}
\includegraphics[width=\linewidth]{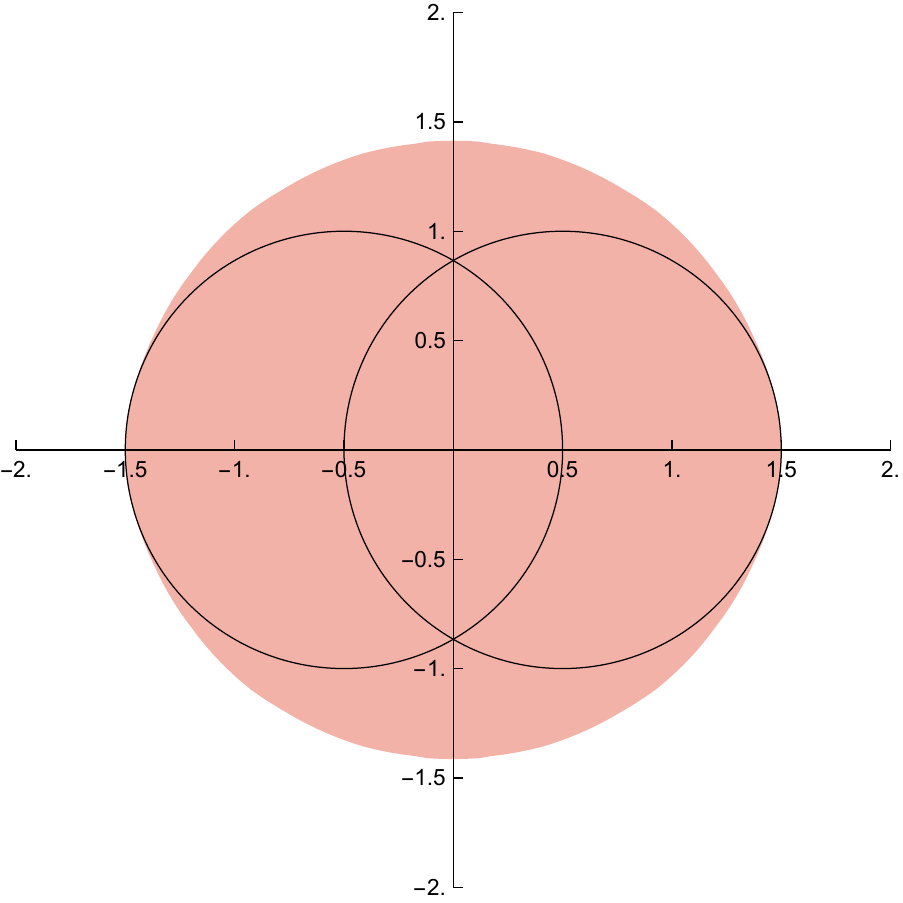}
\end{subfigure}
\caption{The domains of influence of a curvature singularity on the tetrahedron (left) and the octahedron (right). 
}
\label{fig:domain}
\end{figure}

This leads to the notion of the \textit{domain of influence} of a curvature singularity with angle $\alpha$, defined as the
open two-dimensional region, including and surrounding a circle pair $(S_{p}^\delta,S_{p'}^\delta)$, which consists of all points
where a singularity of strength $\alpha$ would cause $\bar{d}/\delta$ to deviate from its flat-space value.
Fig.\ \ref{fig:domain} illustrates the situation for $\alpha\! =\! \pi$ and $\alpha\! =\! 2\pi/3$, associated with the deficit angles
on the tetrahedral and octahedral surfaces respectively. The reason why a curvature singularity can
influence the average sphere distance, even when it lies outside either of the circles, is twofold. Firstly, a 
geodesic between a pair $(q,q')$ of points from the two circles can of course pass through a region that lies outside the circles
and be influenced by a singularity located there. Secondly, because of the presence of the singularity, geodesics between points
will not always be unique, but can occur in pairs that pass on either side of the singularity. This can lead to a geodesic shortcut between
given points $q$ and $q'$, compared with the situation in flat space.

The relevant quantity we need to determine for our purposes is the maximal diameter of the domain of influence for a
given angle $\alpha$. When measuring the curvature profile with the method described above, we must make sure that this diameter
does not exceed the edge length $L$ of the given polyhedron, which is the same as the distance between neighbouring singularities.
If we call ${\cal D}(\alpha)$ the maximal diameter in units of $\delta$, it follows that the largest allowed circle size is 
$\delta_{\rm max}\! =\! L/{\cal D}(\alpha)$. The maximal diameter for the tetrahedral surface 
can be determined from geometric considerations and is given by ${\cal D}(\pi)\! =\! \sqrt{9+4 \sqrt{2}}\! \approx\! 3.828$, corresponding
to the vertical axis in the left diagram of Fig.\ \ref{fig:domain}. As the deficit angle $\alpha$ decreases towards $2\pi/3$, the domain of
influence shrinks along both axes until it reaches the left- and rightmost points of the double circle. The domain keeps shrinking along the
vertical axis to a value below 3 (Fig.\ \ref{fig:domain}, right). It cannot shrink further along the horizontal direction since a singularity 
inside one (or both) of the circles always leads to a nonflat result for $\bar{d}/\delta$. We conclude that for the octahedron and
the remaining platonic solids with $\alpha\! =\! \pi/2$, $\pi/3$ and $\pi/5$, we have ${\cal D}(\alpha)\! =\! 3$.

\section{Measuring curvature profiles}
\label{sec:measure}

The insights gained above will now be applied to evaluate curvature profiles for the surfaces of the five convex regular 
polyhedra depicted in Table \ref{tab:platonic}. Each surface consists of identical polygons, whose edges have all the same 
length. Its Gaussian curvature is distributed equally over all vertices, which implies that each deficit angle has
size $\alpha\! =\! 4\pi/\#$vertices. Since we are interested in the effectsof how the same amount of curvature is distributed
over a given spatial volume,
we compare the results for surfaces of the same area, which we take to be $4\pi$, the area of a two-sphere
of unit radius. The resulting values for the edge lengths $L$ and the associated maximal circle radii $\delta_{\rm max}$
can be found in Table \ref{tab:platonic}.
\begin{center}
\newcommand\ph{0.09}
\begin{table}[ht]
\small
  \begin{tabular}{ | c | c | c | c | c | c | c | c |}
    \hline
 Platonic solid    &    surface &    \begin{tabular}[c]{@{}c@{}}elementary\\region\end{tabular} &\parbox{1.2cm}{\hspace{0.5cm}$\#$ \\vertices} & 
   \parbox{0.9cm}{\hspace{0.3cm}$\#$ \\edges}  &   \parbox{0.9cm}{\hspace{0.3cm}$\#$ \\faces}  
& $L$   &  $\delta_{\rm max}$ \\ \hline
tetrahedron & \raisebox{-0.4 \height}{\includegraphics[height=\ph\textwidth]{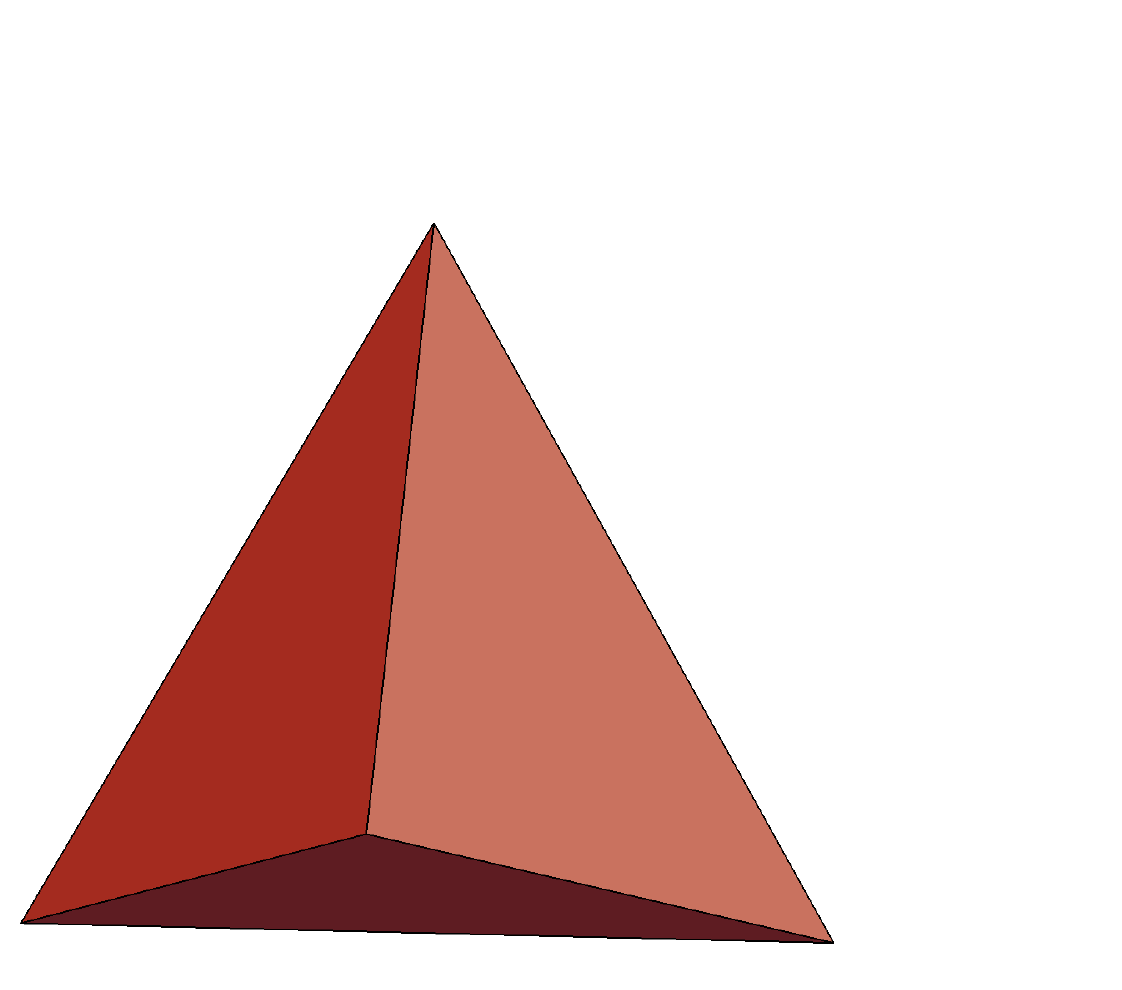}}&
\raisebox{-0.4 \height}{ \includegraphics[height=\ph\textwidth]{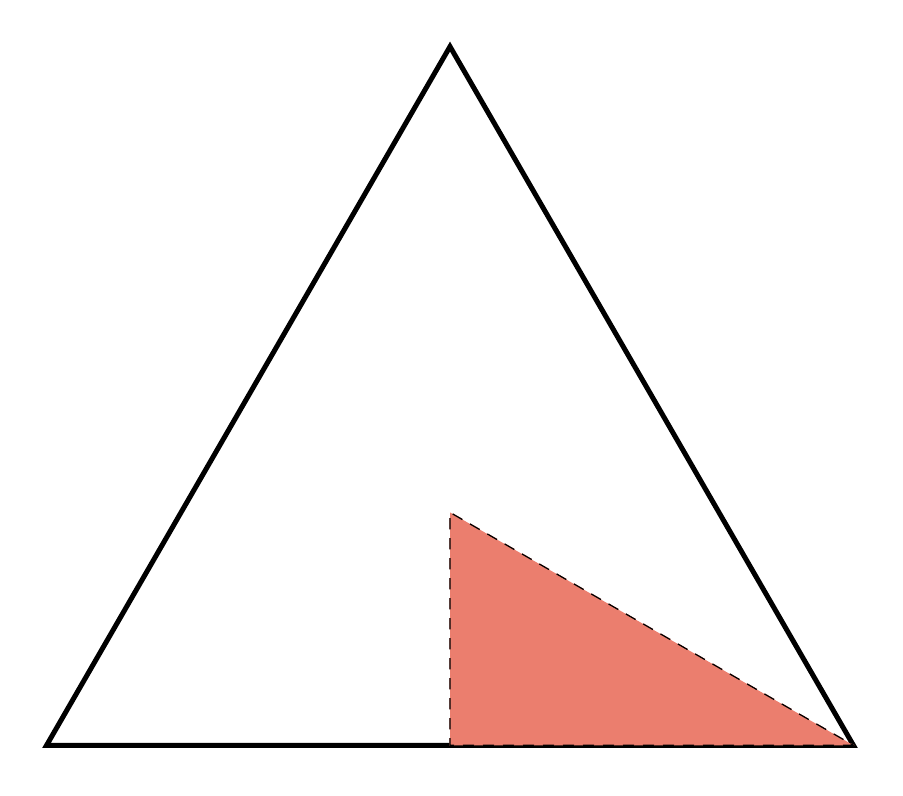}} &4&6&4& 2.694&0.7036 \\ \hline
octahedron & \raisebox{-0.4 \height}{ \includegraphics[height=\ph\textwidth]{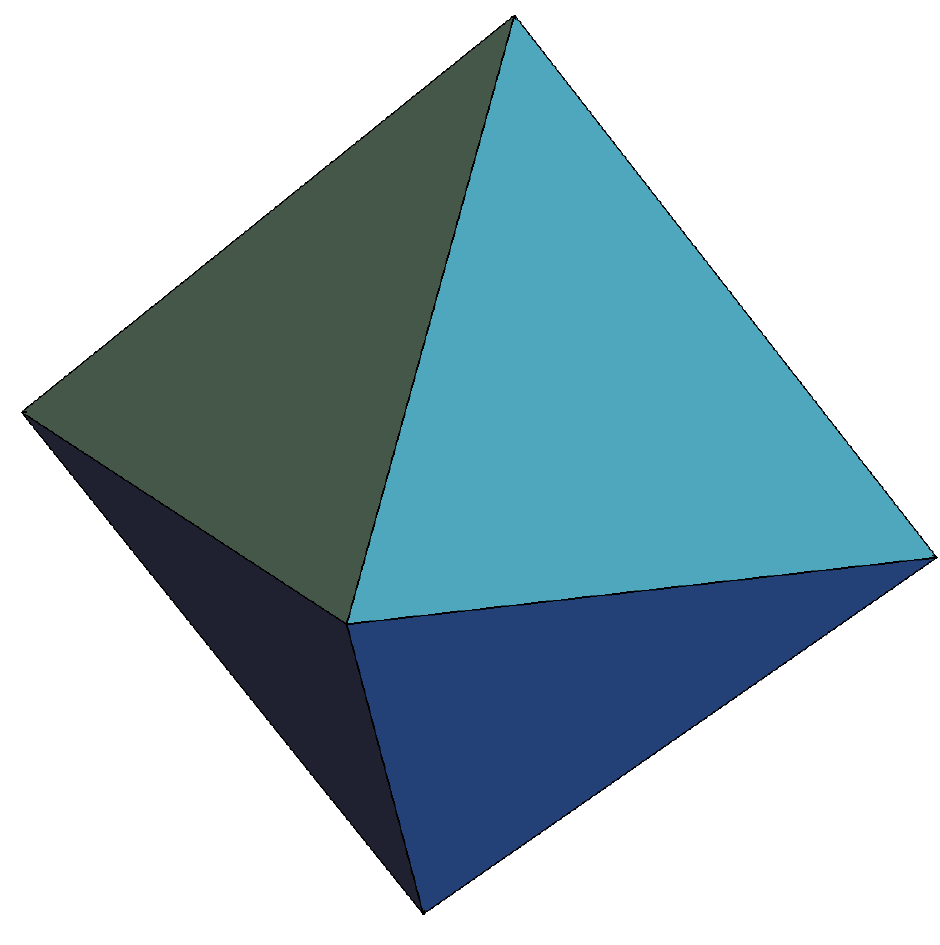} }& 
\raisebox{-0.4 \height}{ \includegraphics[height=\ph\textwidth]{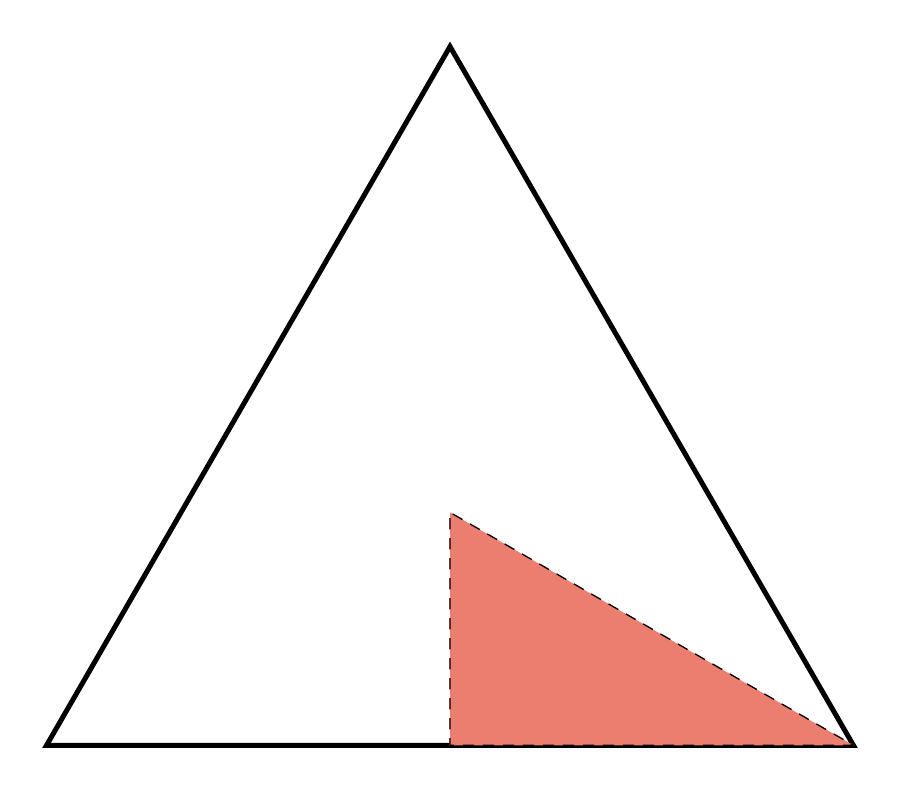}} &6&12&8& 1.905 &0.6349 \\ \hline
 cube &\raisebox{-0.4 \height}{ \includegraphics[height=\ph\textwidth]{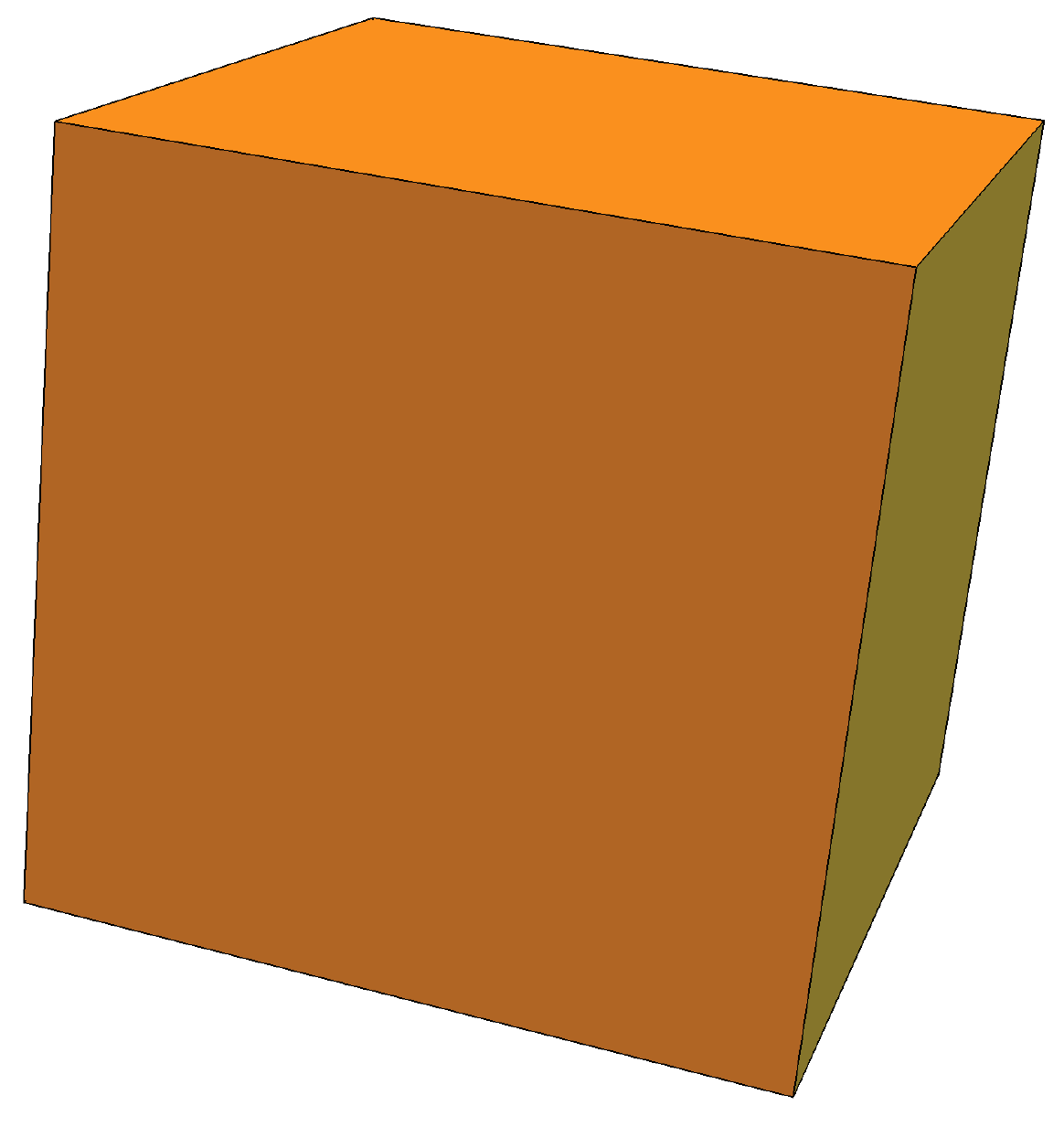} }& 
 \raisebox{-0.4 \height}{ \includegraphics[height=\ph\textwidth]{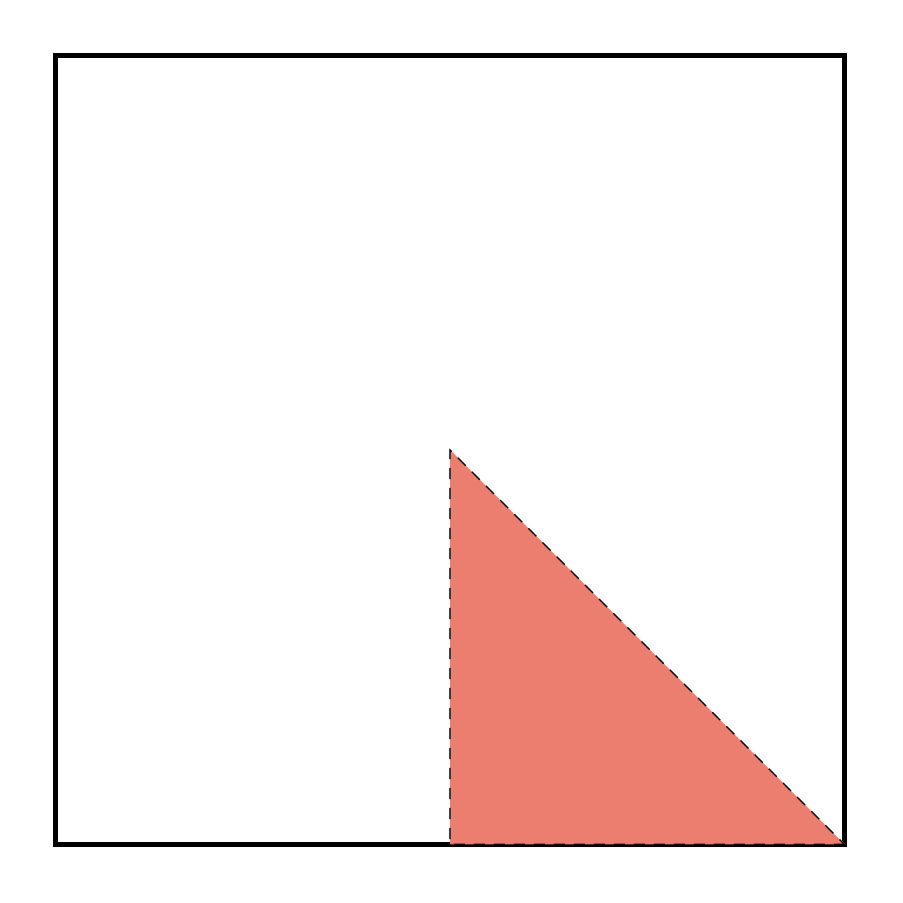} }   &8&12&6& 1.447&0.4824 \\ \hline
 icosahedron &   \raisebox{-0.4 \height}{ \includegraphics[height=\ph\textwidth]{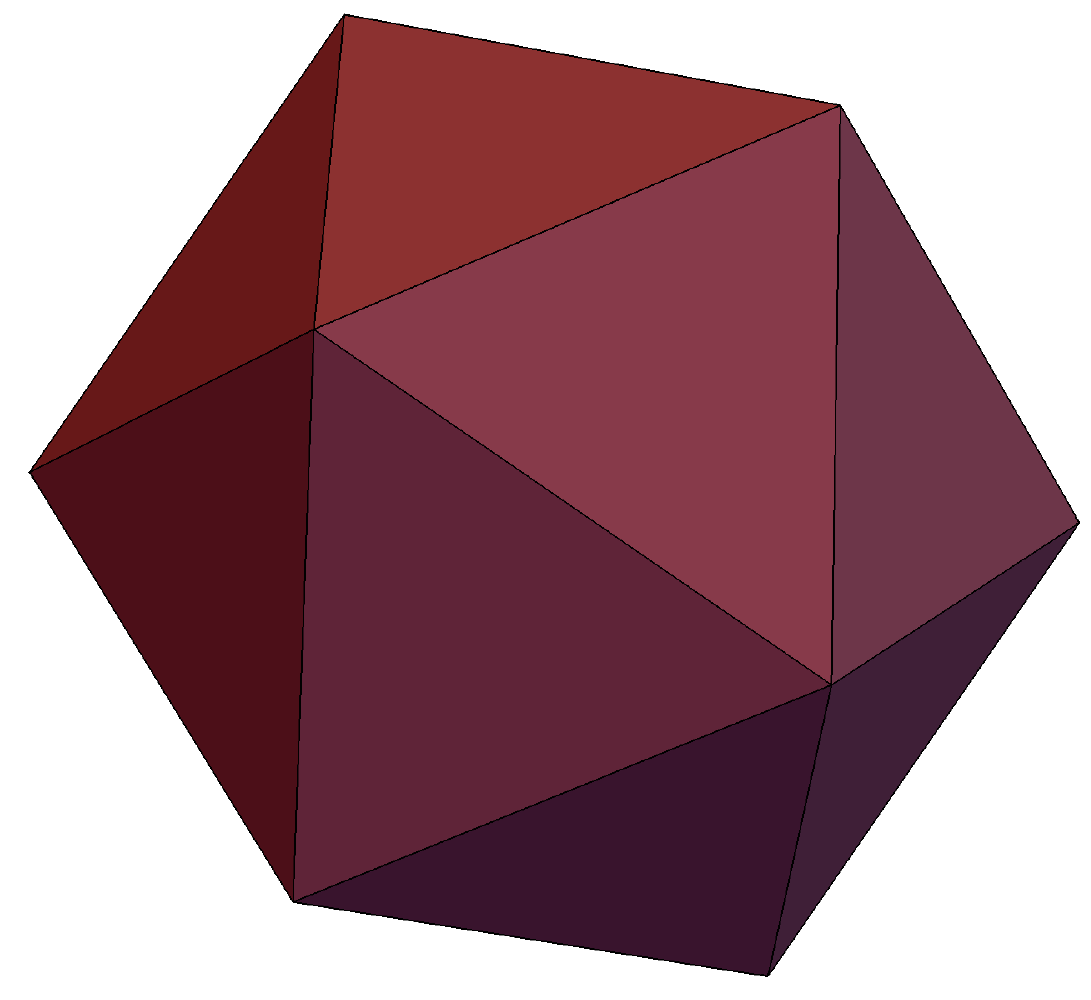} }  & 
 \raisebox{-0.4 \height}{ \includegraphics[height=\ph\textwidth]{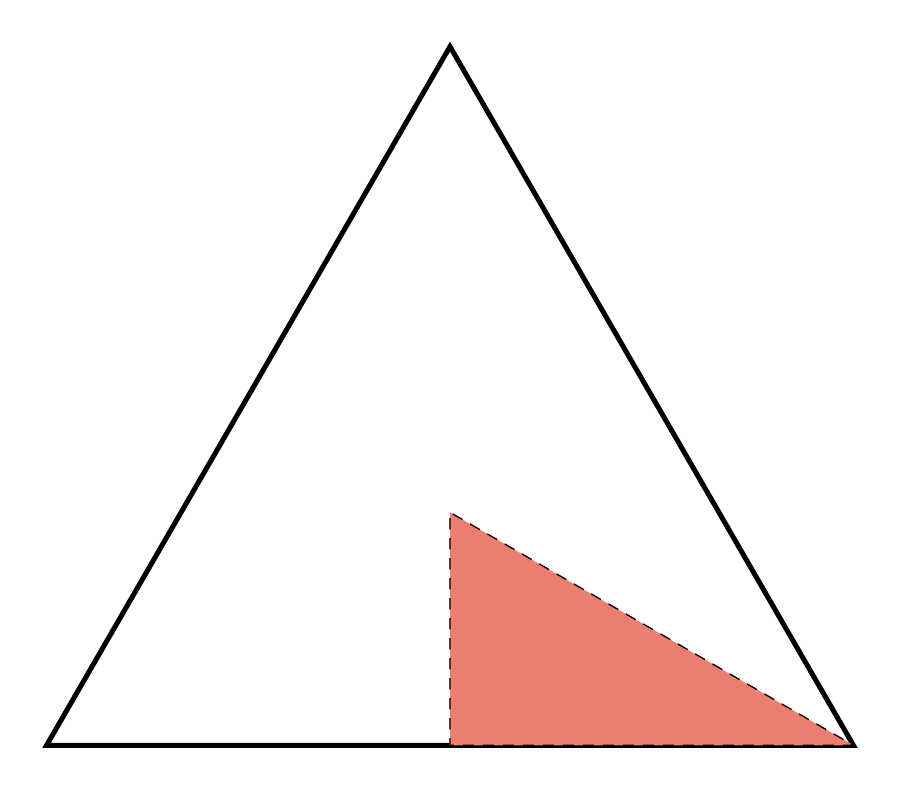} }   &12&30&20& 1.205&0.4015 \\ \hline
 dodecahedron  &  \raisebox{-0.4 \height}{  \includegraphics[height=\ph\textwidth]{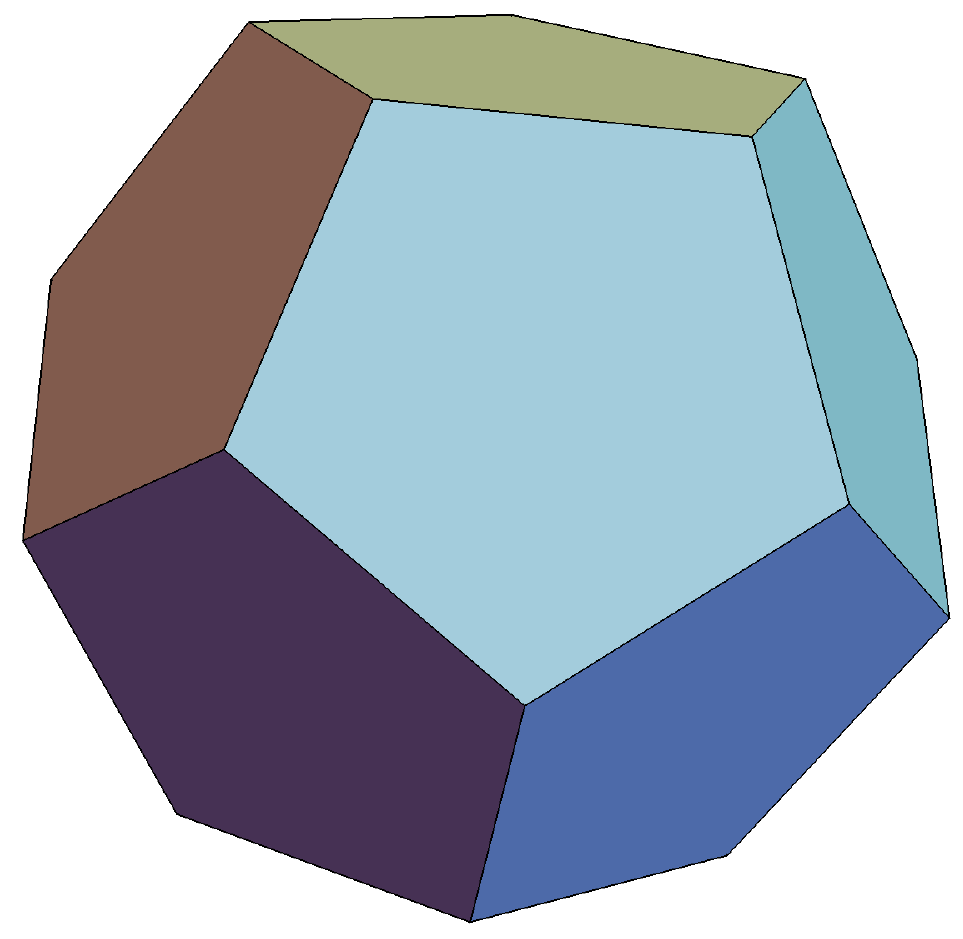} }& 
 \raisebox{-0.4 \height}{   \includegraphics[height=\ph\textwidth]{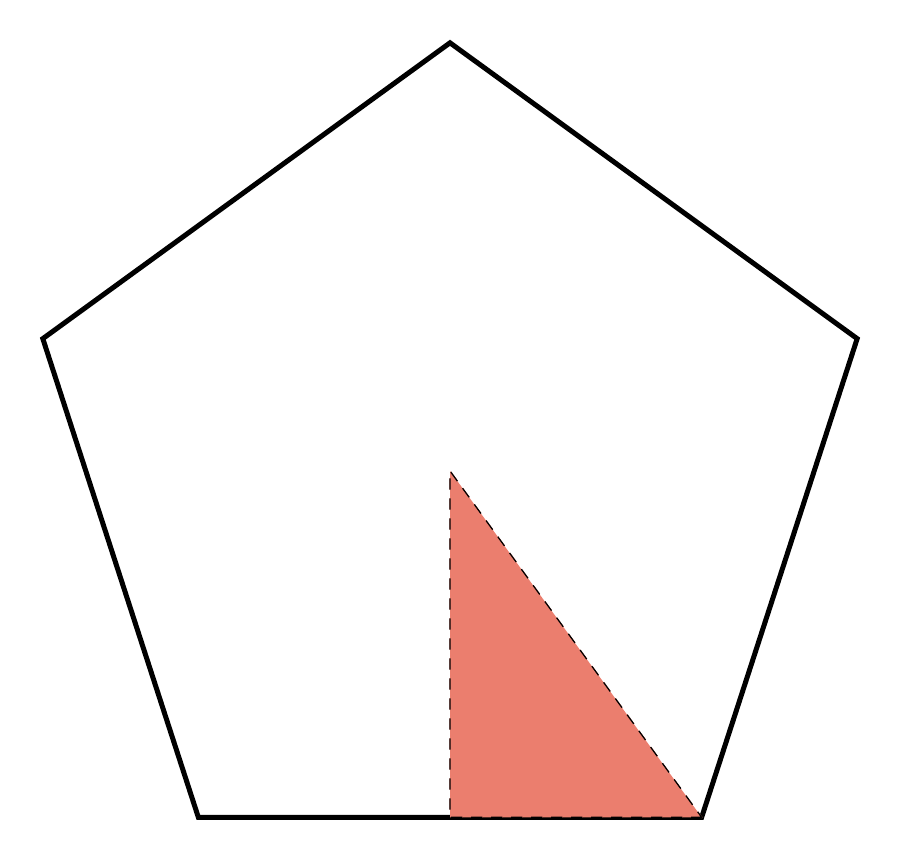} }&20  &30  &12  & 0.780 &0.2601 \\\hline
  \end{tabular}
\caption{Polyhedral surfaces of the Platonic solids, with some of their properties. }
  \label{tab:platonic}
  \normalsize
  \end{table}
\end{center}
\vspace{-0.8cm}
From the limits on $\delta$ it is clear that we will only obtain partial curvature
profiles. It is not straightforward to extend the present method to include a second conical singularity, and
we have not attempted to do so. However, it turns out that for the special case of the tetrahedral surface there is an 
alternative way to extend the $\delta$-range significantly, as will be discussed in Sec.\ \ref{sec:tetra} below. The reason is that one  
can attain a good and computationally explicit control over its geodesics. 

\begin{figure}
\centering
\begin{subfigure}{0.48\textwidth}
\includegraphics[width=\linewidth]{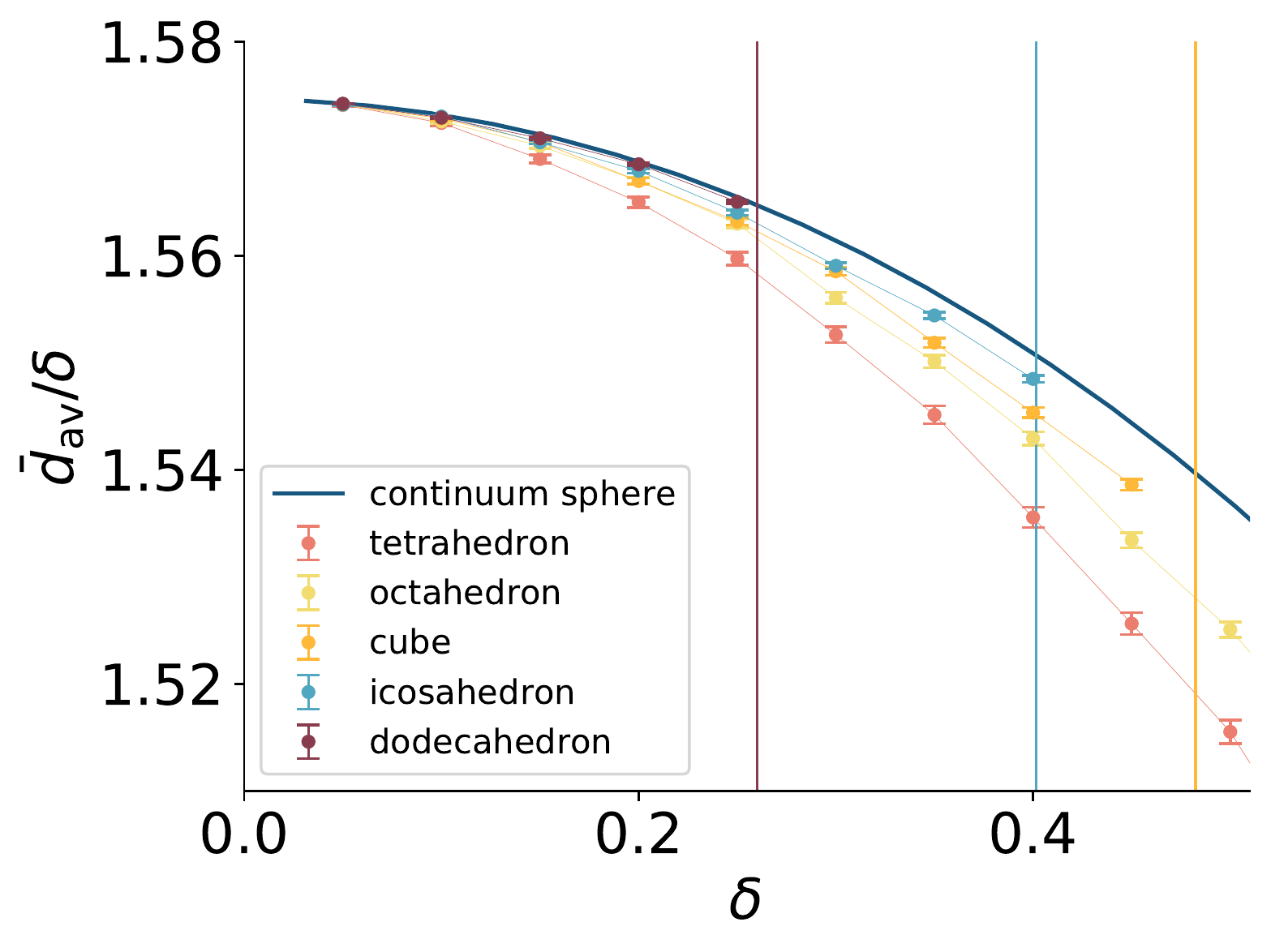}
\label{fig:psr-a}
\end{subfigure}
\hspace{0.02\textwidth}
\begin{subfigure}{0.48\textwidth}
\includegraphics[width=\linewidth]{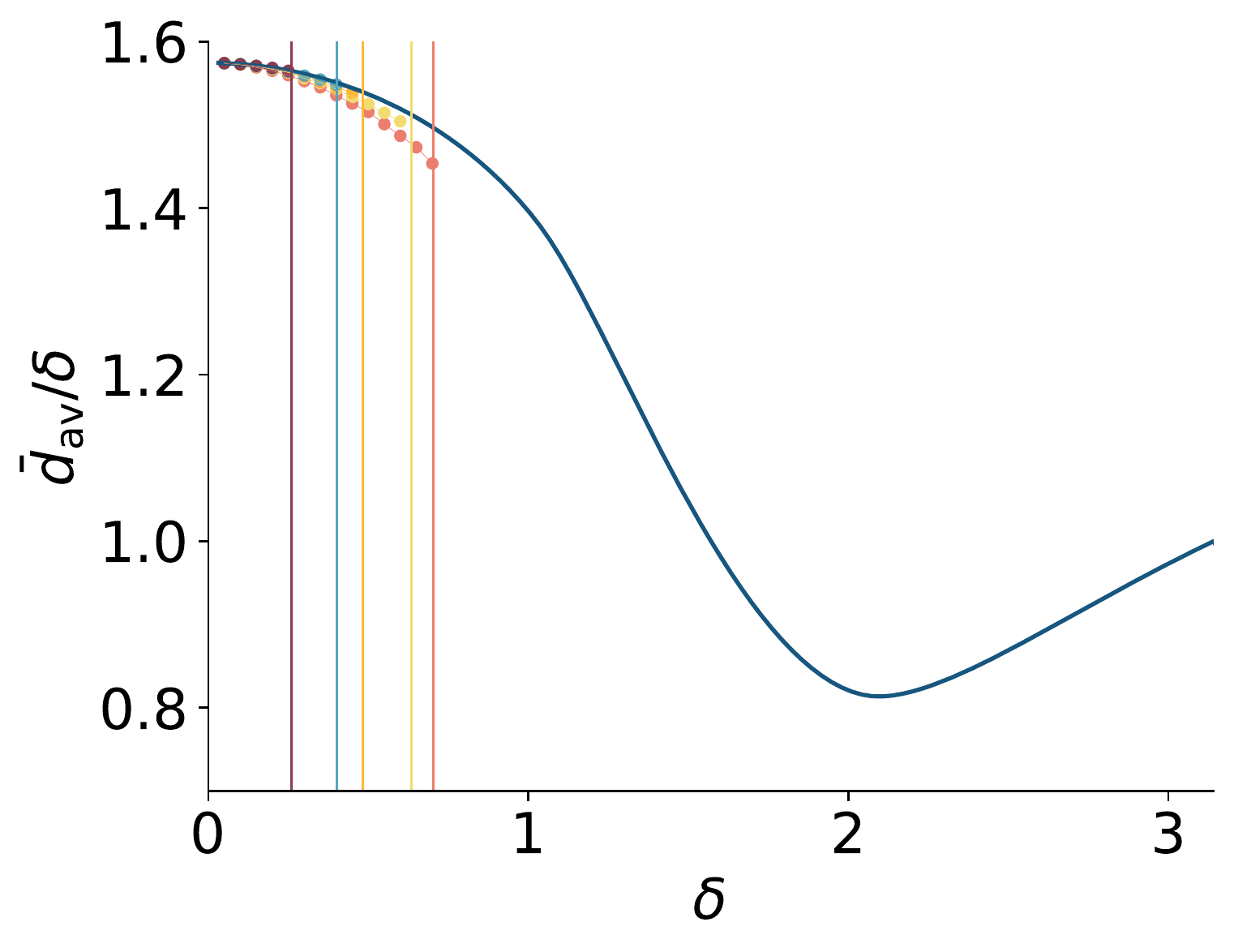}
\label{fig:psr-b}
\end{subfigure}
\caption{Measurements of the curvature profiles $\bar{d}_{\rm av}(\delta)/\delta$ 
of the surfaces of the five Platonic solids, compared to that of the continuum sphere of the same
area (top curve), plotted in the $\delta$-ranges $[0,0.5]$ (left) and $[0,\pi ]$ (right). (Line segments connecting
data points are merely to guide the eye.)}
\label{fig:platonic}
\end{figure}

To compute the spatial average of the average sphere distance on a given polyhedral surface, we use a uniform sampling 
with respect to the flat metric (disregarding the singular points), somewhat similar to what has been done in a quantum context 
\cite{qrc2,qrc3}. Because of the discrete symmetries of the polyhedra, it suffices to pick the first circle centre $p$ of a 
double-circle configuration from what we will call an {\it elementary region}.
This is a triangular subregion of a face, whose corners are given by the midpoint of an edge, one of the endpoints of
the same edge (a singular vertex) and the midpoint of the face, as indicated in Table \ref{tab:platonic}. 
The data for a given value of $\delta\in\, ]0,\delta_{\rm max}]$ are collected as follows:

\vspace{0.15cm}
\begin{enumerate}
\item Sprinkle $n$ points $p_{i}$, $i\! =\! 1,\dots,n$, randomly and uniformly into an elementary region.
\item Construct geodesic circles $S^\delta_{p_{i}}$, using the parametrizations found in Sec.\ \ref{subsec:sd}.
\item For every $p_{i}$, pick a point $p'_{i}$ on $S^\delta_{p_{i}}$ uniformly at random, and construct the geodesic
circle $S^\delta_{p'_{i}}$ around it. 
\item Compute $\bar{d}(S_{p_{i}}^\delta,S_{p'_{i}}^\delta )$ for all pairs $(p_{i}, p'_{i})$. 
\item Average over the results to obtain $\bar{d}_{\rm av}(\delta)$.
\end{enumerate}
\vspace{0.15cm}

\noindent We have collected 10.000 measurements for each data point, with a step size of $\delta\! =\! 0.05$.
The results for the curvature profiles $\bar{d}_{\rm av}/\delta$ 
for the five regular polyhedra are shown in Fig.\ \ref{fig:platonic}, together with the curvature 
profile of the continuum two-sphere for comparison. On the left, we show the behaviour at small distances $\delta\! \lesssim \! 0.5$.
It is clearly distinct for the five spaces, but qualitatively similar. 
(The vertical lines indicate the cutoff values $\delta_{\rm max}$ for the various cases.) 
The data for the dodecahedron resemble those of the continuum sphere most closely, at least within the limited $\delta$-range
considered. Generally speaking, the larger the number of vertices over which the deficit angles are distributed, 
the closer is the match with the sphere curve. Note also that all curves seem to converge to the flat-space value as $\delta\rightarrow 0$,
as one would expect. 

In Fig.\ \ref{fig:platonic}, right, we have zoomed out for a more global comparison with the curvature
profile of the sphere in the range $\delta\in [0,\pi]$. As mentioned above, we will in the following section introduce a different method
to determine geodesic circles on the tetrahedron, which will allow us extend its curvature profile over most of the range depicted here.
Based on the limited data presented here, we conclude that the averaged quantum Ricci curvature of all the investigated spaces is positive,
which is not surprising. 
It is largest for the tetrahedron, for which the decrease of the curve for $\bar{d}_{\rm av}(\delta)/\delta$ is steepest. 

\begin{figure}[t]
\centering
\begin{subfigure}[t]{0.45\textwidth}
\centering
\includegraphics[height=0.7\linewidth]{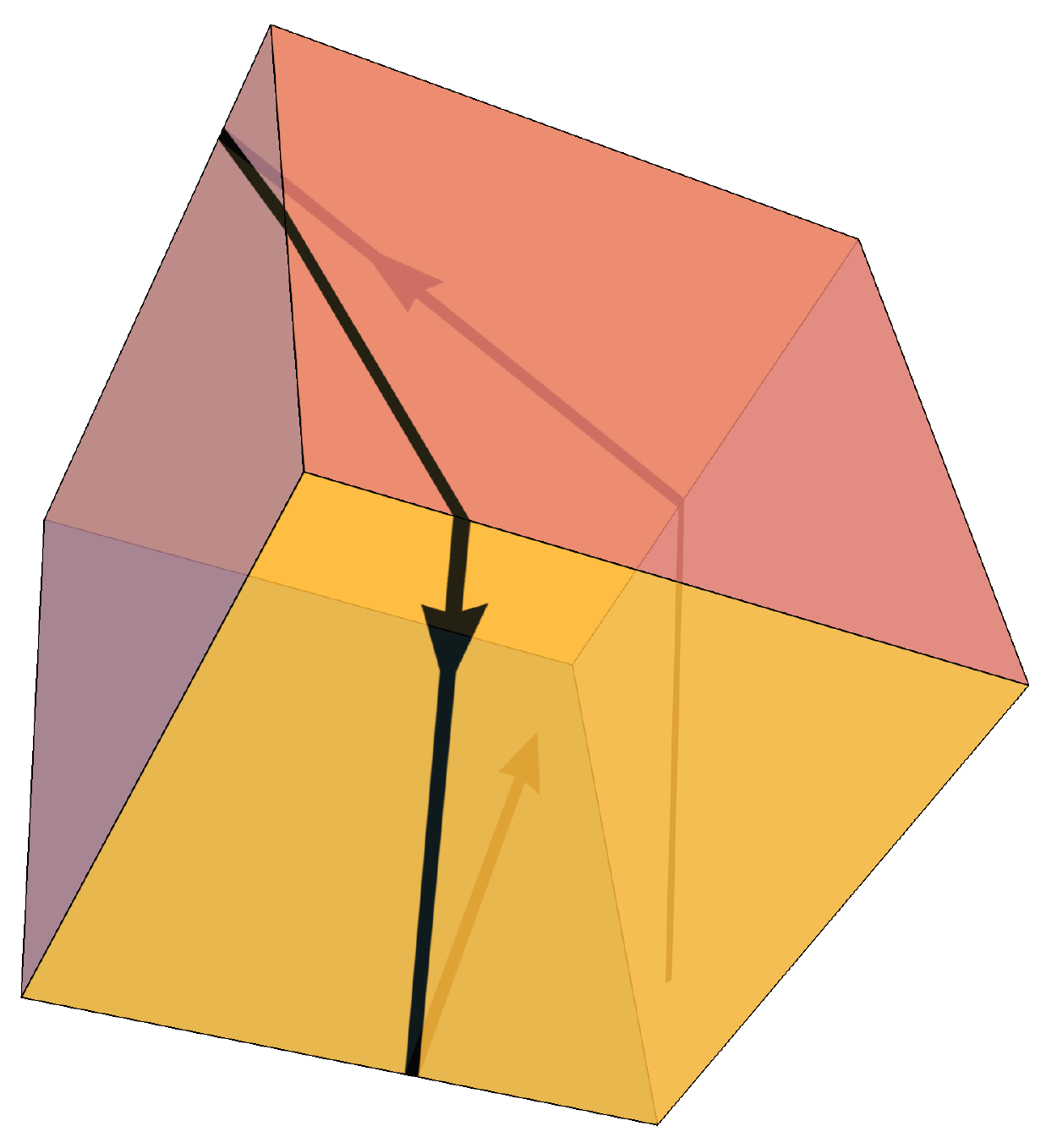}
\end{subfigure}
\hspace{0.02\textwidth}
\begin{subfigure}[t]{0.45\textwidth}
\centering
\includegraphics[height=0.7\linewidth]{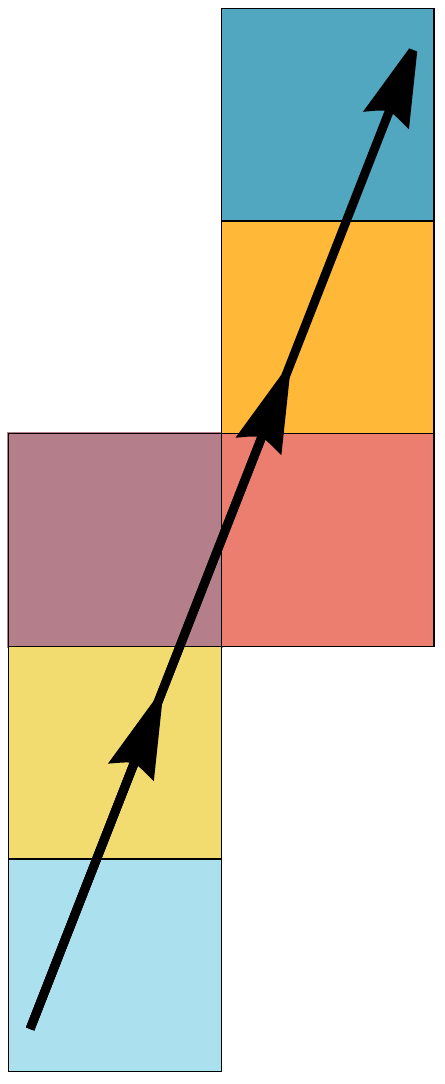}
\end{subfigure}
\caption{Geodesic on a cube (left) and its corresponding development in the plane (right).}
\label{fig:develop-cube}
\end{figure}

\section{Unfolding the tetrahedron}
\label{sec:tetra}

The study of geodesics on regular convex polyhedra, and the classification of closed geodesics in particular, 
is an active area of mathematical research (for example, see \cite{fuchs1,fuchs2}). 
Our primary interest are shortest geodesics between arbitrary pairs of points, a problem that is most straightforwardly tackled for the
tetrahedron. This is fortunate, since the tetrahedron is the most interesting case from our point of view. Of all the Platonic solids, its curvature is 
distributed least homogeneously, while the results on the curvature profiles of the previous section suggest that all other cases lie 
in between those of the tetrahedron and the smooth sphere. 

A general technique one can use to study geodesics on regular polyhedra is by ``rolling"  or ``unfolding" the polyhedron onto the flat two-dimensional
plane along a given geodesic. The so-called development of a polyhedron along a geodesic pro\-ceeds as follows \cite{fuchs1}: 
starting from an arbitrary point $p$ contained in some face $F_0$ of the polyhedral surface and an initial direction, one follows the 
corresponding geodesic (straight line) in $F_0$ until it hits an edge to a neighbouring face $F_1$. Since all faces are flat, the adjacent pair of 
$F_0$ and $F_1$ can in an isometric way be put down in the plane. Following the geodesic as it passes through $F_1$, it will meet another 
edge to some neighbouring face $F_2$, which likewise can be folded out into the plane, and so forth.
The result of this development is a contiguous chain of faces in the plane, which is traversed by the geodesic, taking the form of a straight line.
Fig.\ \ref{fig:develop-cube} illustrates the procedure for a geodesic running along the surface of a cube. The six sides of the cube have a 
different colour coding and will in general appear repeatedly along a given geodesic.
\begin{figure}[t]
\centering
\begin{subfigure}[t]{0.45\textwidth}
\includegraphics[height=0.8\linewidth]{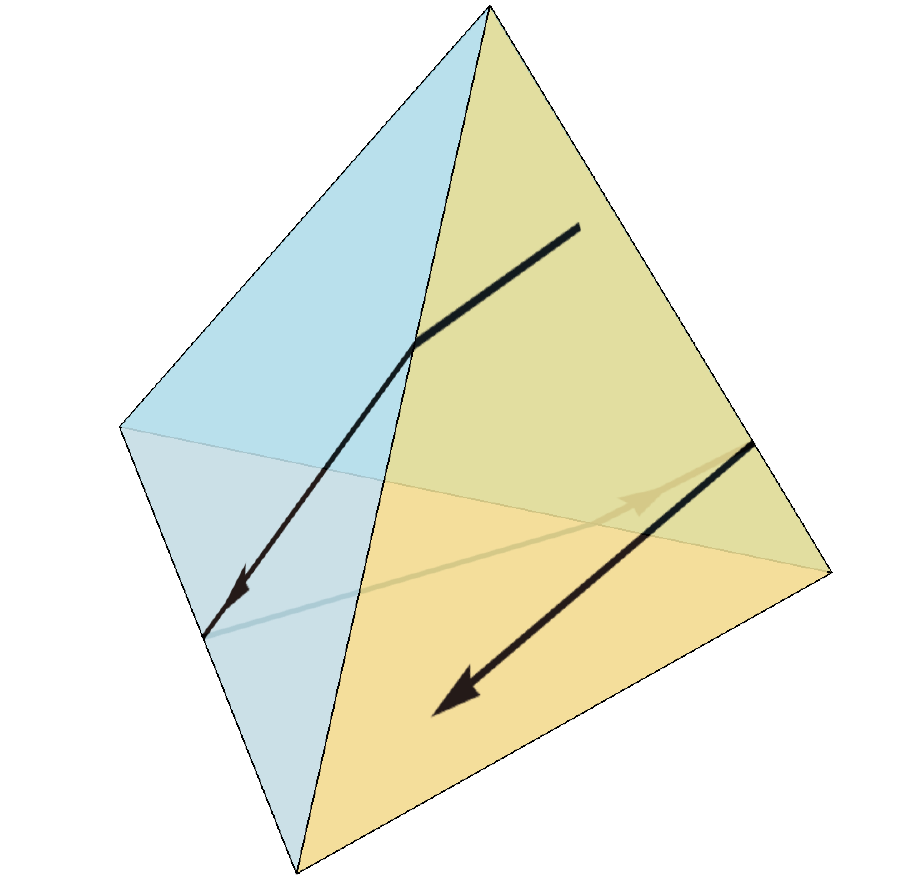}
\end{subfigure}
\hspace{0.02\textwidth}
\begin{subfigure}[t]{0.45\textwidth}
\includegraphics[height=0.8\linewidth]{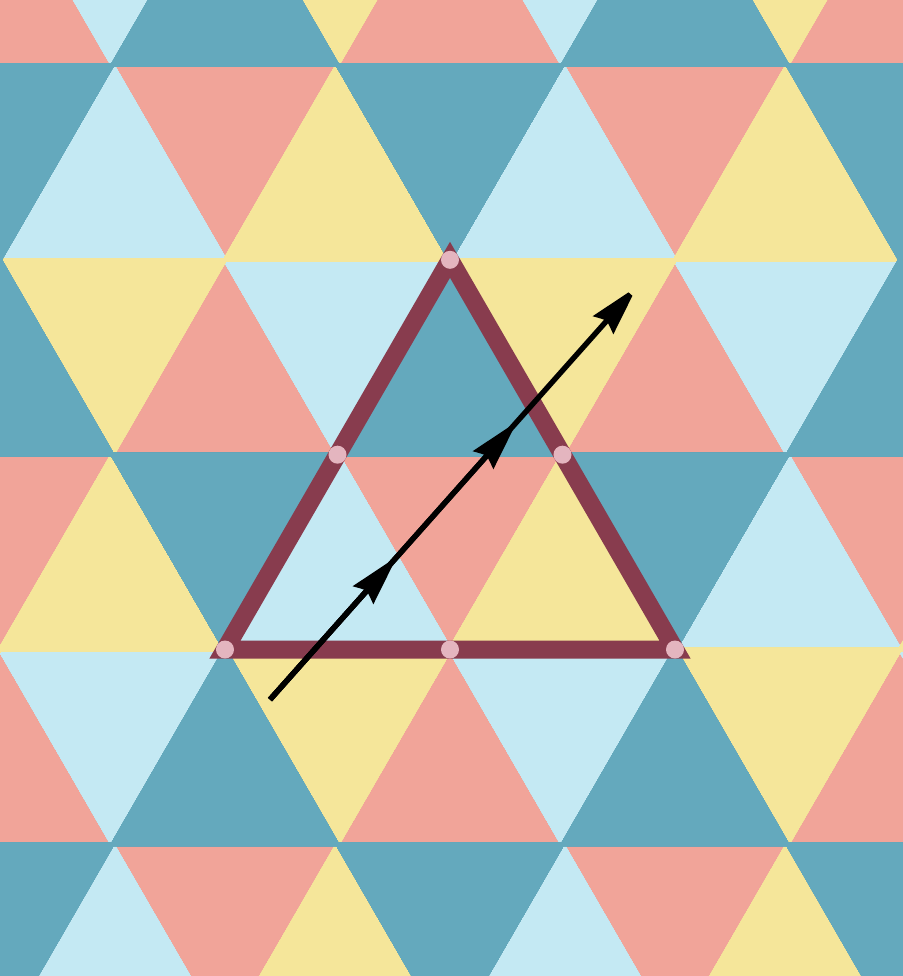}
\end{subfigure}
\caption{Geodesic on a tetrahedron (left) and its development in the plane, which forms part of a regular tiling in terms of the
four faces of the tetrahedron (right). The thick line encloses a fundamental domain $\cal F$.}
\label{fig:develop-tetra}
\end{figure}

The tetrahedron is special among the regular polyhedra in the sense that it has an everywhere consistent development, 
which is independent of the geodesic chosen. The full development in all directions constitutes a regular 
tessellation of the flat plane by 
equilateral triangles, which come in four types or colours, corresponding to the four faces of the tetrahedron. Fig.\ \ref{fig:develop-tetra}
shows an example of a geodesic on the tetrahedron and the corresponding straight line in the regular tiling. 
We have also indicated a triangle-shaped {\it fundamental domain}  $\cal F$, consisting of four triangles that make up a single copy of the tetrahedron. 
The plane can also be thought of as a tessellation by copies of $\cal F$ with alternating orientation, either
with the orientation shown in the figure or an upside-down version rotated by 180 degrees. This planar set-up will allow us to 
construct geodesic circles and compute average sphere distances for radii $\delta$ of up to one tetrahedral edge length, which
is almost four times the range we could cover with the previous method.

\subsection{Geodesic circles on the tetrahedron}
\label{sec:geo}

\begin{figure}[t]
\centering
\begin{subfigure}[t]{0.48\textwidth}
\centering
\includegraphics[height=5.5cm]{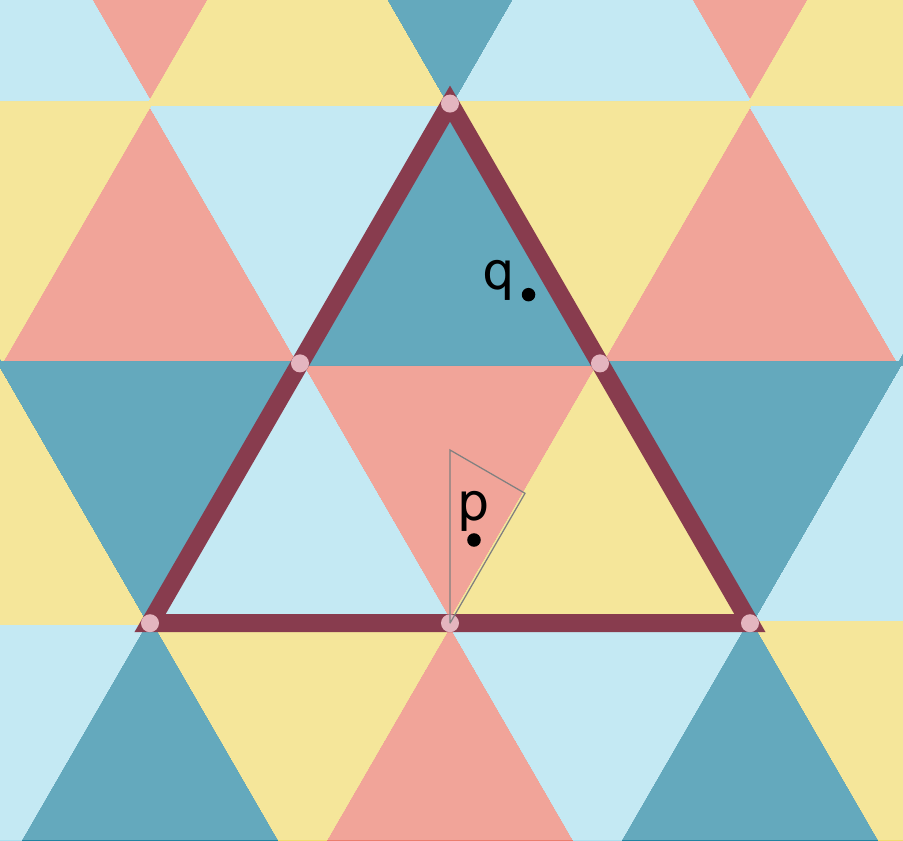}
\end{subfigure}
\hspace{0.02\textwidth}
\begin{subfigure}[t]{0.48\textwidth}
\centering
\includegraphics[height=5.5cm]{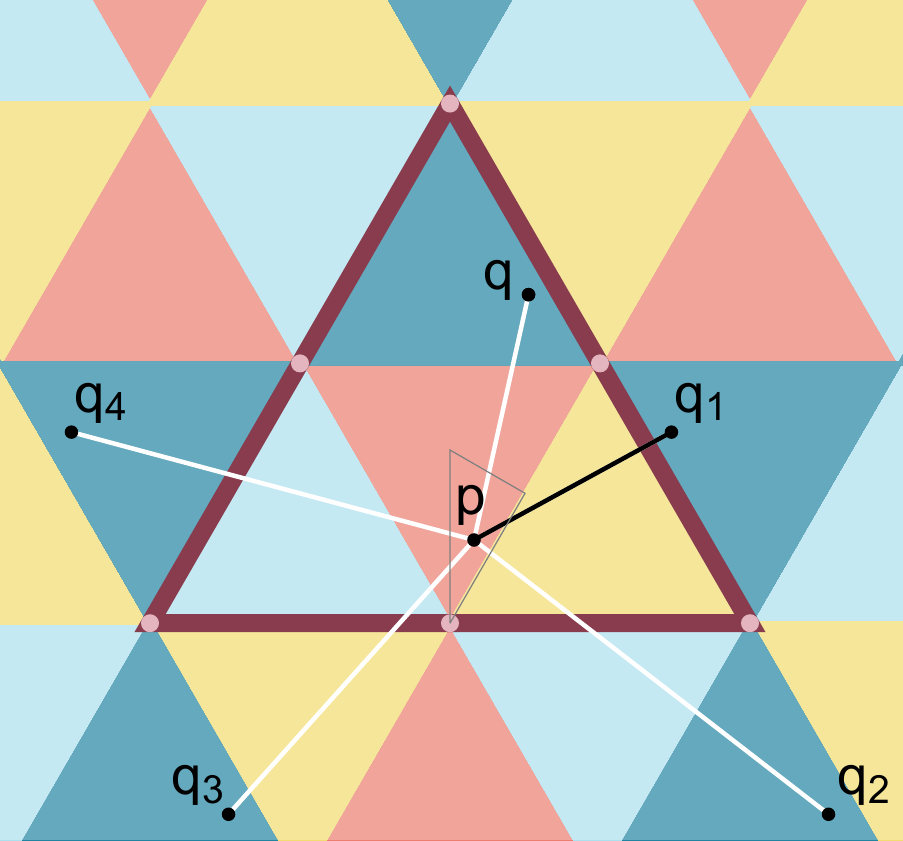}
\end{subfigure}
\caption{Fundamental domain $\cal F$ with a point $p$ in the elementary region of the 
centre triangle of $\cal F$, and a second point $q$ (left). 
To determine the distance between $p$ and $q$, we must consider copies $q_i$ of the point $q$ in neighbouring copies of $\cal F$ (right).
The shortest line is drawn in black. }
\label{fig:tetra}
\end{figure}

The new representation in terms of unfolded tetrahedra allows us relatively easily to determine geodesic circles that enclose more than
one singularity. A more elementary step we need to understand first is how to determine the distance between two arbitrary points
$p$ and $q$ on the surface of the tetrahedron, which is given by the length of the shortest geodesic between them. 
Let us represent the surface by the fundamental domain $\cal F$ of Fig.\ \ref{fig:develop-tetra}. 
Because of the symmetries of the tetrahedron, 
we can without loss of generality choose the point $p$ to lie in (or on the boundary of) the elementary triangular region in the central
triangle of $\cal F$. The second point $q$ can be located anywhere in $\cal F$ (Fig.\ \ref{fig:tetra}, left). 
The key observation now is that the unique straight line 
one can draw from $p$ to $q$ in $\cal F$ clearly corresponds to a geodesic on the tetrahedron, but not necessarily the shortest one.
This can happen because the tetrahedral surface is obtained by appropriate pairwise identifications of the six boundary edges
of the domain $\cal F$, and the shortest geodesic to $q$ may cross one or more of these boundaries. One
can easily find this geodesic by going back to the tessellation of the plane and considering a sufficiently large 
neighbourhood of the fundamental domain $\cal F$, consisting of $\cal F$ itself and neighbouring
copies ${\cal F}_i$, $i\! =\! 1,2,\dots$, of $\cal F$. On each of these copies, we mark the unique copy 
$q_i$ of the point $q$ (Fig.\ \ref{fig:tetra}, right). 
The length of the shortest straight line between $p$ and either $q$ or any of the $q_i$ is the searched-for geodesic distance
between $p$ and $q$. 

This way of representing the tetrahedron and its geodesics is not unlike our previous representation of the
geometry of an isolated conical singularity in terms of a plane with a wedge removed: the nature of the geodesics is
maximally simple (straight lines), but one pays a price in the form of nontrivial identifications.

Building on the above insight on how geodesic distances are determined, we next discuss the nature of geodesic 
circles of radius $\delta$
based at a point $p$.
It is best illustrated by referring to the convex set ${\cal C}_p$ 
of all points $q$ {\it in the tessellated plane}, for given $p$ in the elementary region, which are 
closer to $p$ than any of their copies (in $\cal F$ or any of the ${\cal F}_i$). To construct ${\cal C}_p$, one needs to recall the status of vertices
in the tessellated plane. On the original tetrahedron, they corresponded to singularities with a deficit angle $\alpha\! =\! \pi$. 
Cutting open the tetrahedron along three of its edges and putting it in the plane results in a copy of the fundamental domain $\cal F$, which
can be thought of as a region in the plane with three such wedges removed. Since all wedges have an angle $\pi$, one
fundamental domain can be glued into another's ``missing wedge". Repeating this gluing throughout the plane leads to a tiling of the plane 
which preserves all neighbourhood relations between pairs of tetrahedral faces sharing a common edge. 
What is not preserved is the number of faces meeting at a given vertex, which is three on the original tetrahedron and six
in the plane.
\begin{figure}
\centering
\includegraphics[height=6cm]{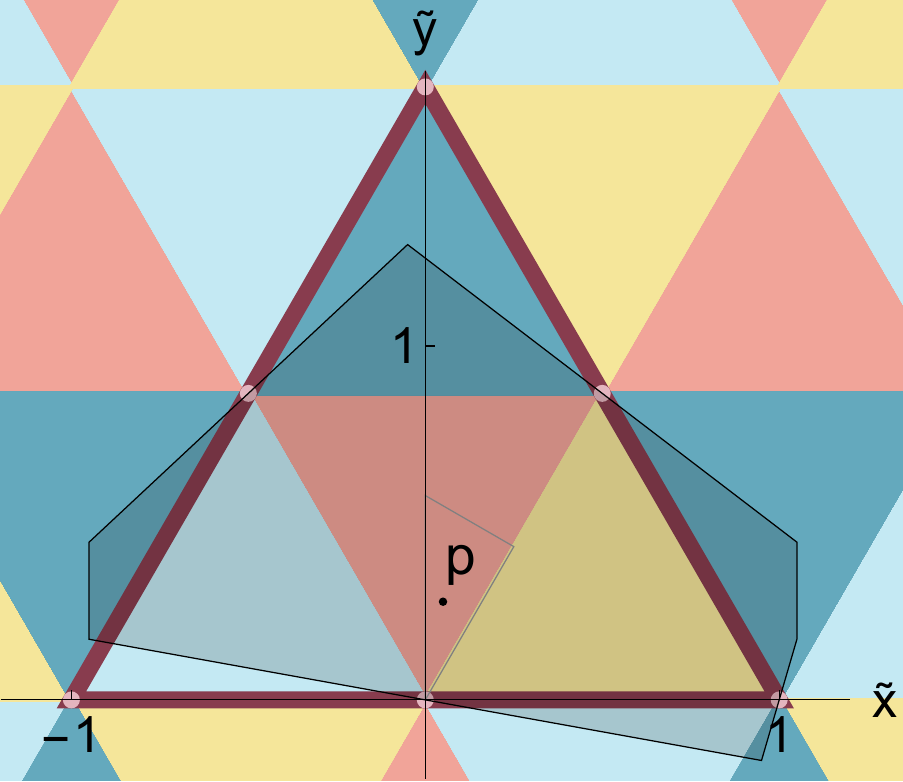}
\caption{The convex polygon ${\cal C}_p$ superimposed on the tessellated plane. The axes show the orthonormal coordinate system $(\tilde{x}, \tilde{y})$ used to construct the region.}
\label{fig:tetra-cp}
\end{figure}

The region ${\cal C}_p$ in the tessellated plane can be constructed from symmetry considerations. 
For a general point $p$ inside the elementary
region, its boundary is a convex polygon
with six corner points $S_i$, $i\! =\! 1,\dots,6$. Its sides are formed by four straight line segments through the four vertices closest
to $p$, where each line segment is perpendicular to the line from $p$ to the vertex in question. 
In addition, there are two parallel lines (vertical lines in Fig.\ \ref{fig:tetra-cp}) equidistant from $p$, which
get mapped onto each other by one of the translation symmetries of the tessellated plane. 
Note that by construction the area of ${\cal C}_p$ is the same as the surface area of the tetrahedron.

To arrive at a quantitative description, we introduce an orthonormal coordinate system $(\tilde{x},\tilde{y})$ in the plane,
whose origin $(0,0)$ is the midpoint between the two bottom edges of the fundamental domain $\cal F$ (Fig.\ \ref{fig:tetra-cp}).
For definiteness, let us assume that all edges have unit length, $L\! =\! 1$.
Note that the elementary triangular region, which by assumption contains the point $p\! =\! (x,y)$, 
is spanned by the three corner points $(0,0)$, $(0,1/\sqrt{3})$ and $(1/4,\sqrt{3}/4)$, which means that the ranges of
$x$ and $y$ are limited to $x\!\in\! [0,1/4]$ and $y\!\in\! [0,1/\sqrt{3}]$.
In this coordinate system, 
the coordinates of the corner points of the region ${\cal C}_p$ are given by
\begin{equation}
S_1\! =\! \Big( x-1, \frac{x (1\! -\! x)}{y} \Big),\; 
S_2\! =\!\Big( x-1, \tfrac{1}{2} \big(\sqrt{3}-\frac{1-4 x^2}{\sqrt{3}\! -\! 2y} \big)\! \Big),\;
S_3\! =\!\Big( -x, \tfrac{1}{2} \big(\sqrt{3}+\frac{1-4 x^2}{\sqrt{3}\! -\! 2y} \big)\! \Big),\nonumber 
\end{equation}
\begin{equation}
S_4\! =\!\Big( x+1, \tfrac{1}{2} \big(\sqrt{3}-\frac{1-4 x^2}{\sqrt{3}\! -\! 2y} \big)\! \Big),\;
S_5\! =\! \Big( x+1, \frac{x (1\! -\! x)}{y} \Big),\; 
S_6\! =\! \Big( 1-x, -\frac{x (1\! -\! x)}{y} \Big),\; 
\label{spoints}
\end{equation}
for $y\not= 0$, and by
\begin{equation}
S_1\! =\! (-1,0),\; S_2\! =\! \Big(\! -1,\frac{1}{\sqrt{3}}\Big),\; S_3\! =\! \Big( 0,\frac{2}{\sqrt{3}} \Big),\; 
S_4\! =\!\Big( 1,\frac{1}{\sqrt{3}}\Big),\; S_5\! =\! S_6\! =\! (1,0),
\label{snot}
\end{equation}
for $y\! =\! 0$. In the latter case, corresponding to $p\! =\! (0,0)$, ${\cal C}_p$ becomes a five-cornered polygon that is equal to one half
of a regular hexagon.

We have now set the stage for analyzing geodesic circles $S_p^\delta$ centred at $p$. In what follows, we will exclude
the case $p\! =\! (0,0)$, where the circle centre would coincide with a singularity of the tetrahedron. For added simplicity,
we will also assume that $p$ lies in the interior of the elementary region, which is the generic case. An analogous
treatment of points along the boundary is completely straightforward. 

Since in an open 
neighbourhood of $p$ space is flat and Euclidean, for sufficiently small $\delta$ the set of points $S_p^\delta$ is
an ordinary circle of radius $\delta$ and circumference $2\pi\delta$. 
As we increase $\delta$ continuously, this continues to be the case until the circle
reaches the singularity closest to $p$, located at $(\tilde{x},\tilde{y})\! =\! (0,0)$, where it also meets the line segment that
forms part of the boundary of ${\cal C}_p$. The corresponding circle radius is $\delta\! =\! \sqrt{x^2+y^2}$. 
As we increase $\delta$ further, we need to take into account the nontrivial deficit angle associated with the point
$(0,0)$. It implies that the piece of line segment to one side of the vertex $(0,0)$ should be identified with the piece to the other side.
For the circle of radius $\delta$ we draw in the tessellated plane, it means that the circle segment beyond this line segment
is {\it not} part of $S_p^\delta$. The ``circle" $S_p^\delta$ with radius $\delta\! > \! \sqrt{x^2+y^2}$ therefore has a length
that is shorter than $2\pi\delta$. The story repeats itself when we increase $\delta$ further and $S_p^\delta$ meets the 
singularity at $(\tilde{x},\tilde{y})\! =\! (1/2,\sqrt{3}/2)$, which is second-closest to $p$. From this radius onwards, a second
circle segment will be removed from the ``na\"ive" circle we can draw around $p$ in the plane. The same happens when $S_p^\delta$
reaches and passes the remaining two singularities, at $(-1/2,\sqrt{3}/2)$ and $(-1,0)$ respectively. 

Since the tetrahedron has a finite diameter, the size of $S_p^\delta$ must be zero above some maximal value of $\delta$,
given by the distance of the point(s) furthest away from the given point $p$, the antipode(s) of $p$.\footnote{The antipode is a
single point on the tetrahedron, but appears in multiple images on the boundary of ${\cal C}_p$.} In this context, it should be noted that
the three points $S_1$, $S_5$ and $S_6$ of eq.\ (\ref{spoints}) are equidistant to $p$, with distance $d_1$, as are the three points 
$S_2$, $S_3$ and $S_4$, with distance $d_2$. Which of the two distances is larger and therefore sets the upper limit for the radius $\delta$ 
depends on the location of $p$ in the elementary region. It turns out that the extremal cases of the closest and the furthest
antipodes are associated with points $p$ on the boundary of the elementary region. The antipode is closest when $p$ lies
in the middle of an edge, $p\! =\! (1/4,\sqrt{3}/4)$, in which case the antipode has distance $d_1\! =\! d_2\! =\! 1$ and also
lies in the middle of an edge. By contrast, the point $p$ with the most distant antipode, with $d_1\! =\! d_2\! =\! 2/\sqrt{3} \approx \!
1.155$, lies at the centre of a face, with $p\! =\! (0,1/\sqrt{3})$. 

\begin{figure}[t]
\includegraphics[height=0.41\linewidth]{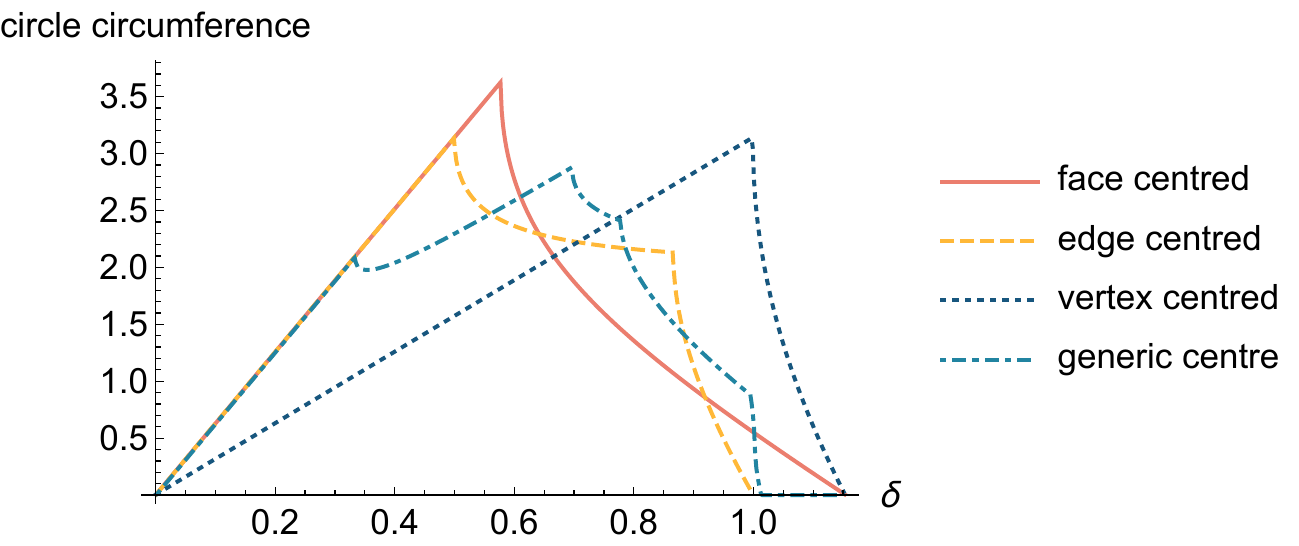}
\caption{The circumference of geodesic circles $S_p^\delta$ on the tetrahedron as a function of the radius $\delta$, both 
in units of edge length, for various locations of the centre $p$.}
\label{fig:tetra-circ}
\end{figure}

These results are illustrated in Fig.\ \ref{fig:tetra-circ}, which shows the
circumference of $S_p^\delta$ as a function of the geodesic radius $\delta$ for various choices of $p$. 
The computation is done by identifying which arcs of the ``na\"ive" circle at radius $\delta$, parametrized by an
ordinary rotation angle $\theta$, contribute to $S_p^\delta$
and by adding up the arc lengths. For all curves, we see
an initial linear rise, characteristic of a circle in the flat plane, with the exception of $p$ at a vertex, for which the linear
slope is flatter, corresponding to that of a cone with deficit angle $\pi$. For a generic location of the centre $p$, 
there subsequently are four cusps, corresponding to the radii where the circle $S_p^\delta$ meets one of the singularities,
as discussed above. For non-generic $p$ some of the cusps can merge. When $p$ lies at the centre of an edge, it is clear that 
the two closest singularities are reached simultaneously at $\delta\! =\! 0.5$, and the two remaining ones at $\delta\! =\! \sqrt{3}/2\! 
\approx \! 0.866$. In this case, the
antipode has the minimal distance 1 from $p$, which is why the curve in Fig.\ \ref{fig:tetra-circ} stops at $\delta\! =\! 1$. 
When $p$ lies at the centre of a face, the three closest singularities are reached
simultaneously at $\delta\! =\! 1/\sqrt{3}\!\approx \! 0.577$, giving rise to a single cusp. The antipode, which in this case is maximally far away from $p$, 
coincides with the remaining vertex, and the corresponding curve ends at $\delta\! \approx\! 1.155$. The same is true for the
dual case, where the construction starts at a vertex. For comparison, the analogous curve for a smooth two-sphere of the same area as
the tetrahedron would end at $\delta\! \approx\! 1.166$.

The above analysis and Fig.\ \ref{fig:tetra-circ} underline the strong inhomogeneity and aniso\-tro\-py of the tetrahedral surface 
and at the same time fix an upper bound, $\delta\! =\! 1$, up to which the notion of a geodesic circle $S_p^\delta$ is
meaningful for {\it arbitrary} locations of $p$. This is relevant when we start taking spatial averages of average distances between
circles of radius $\delta$ in order to obtain the curvature profile of the tetrahedron, which is the subject of the next section.

\subsection{Computing the curvature profile}

\begin{figure}
\centering
\begin{subfigure}{0.45\textwidth}
\includegraphics[width=0.95\linewidth]{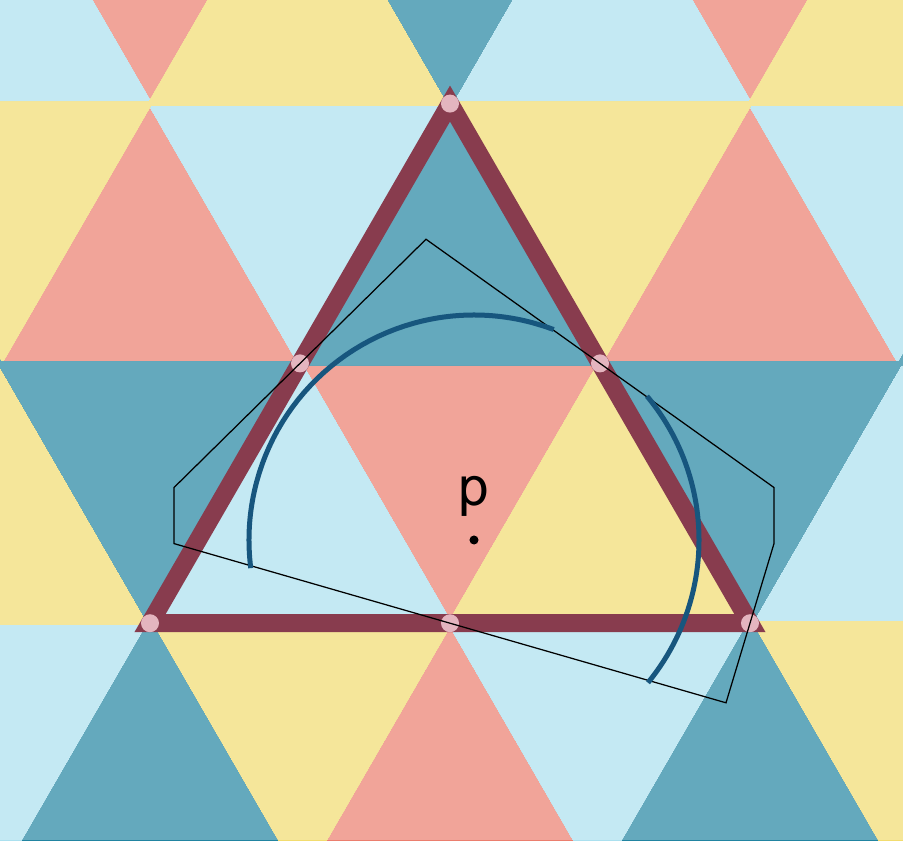}
\end{subfigure}
\hspace{0.02\textwidth}
\begin{subfigure}{0.45\textwidth}
\includegraphics[width=0.95\linewidth]{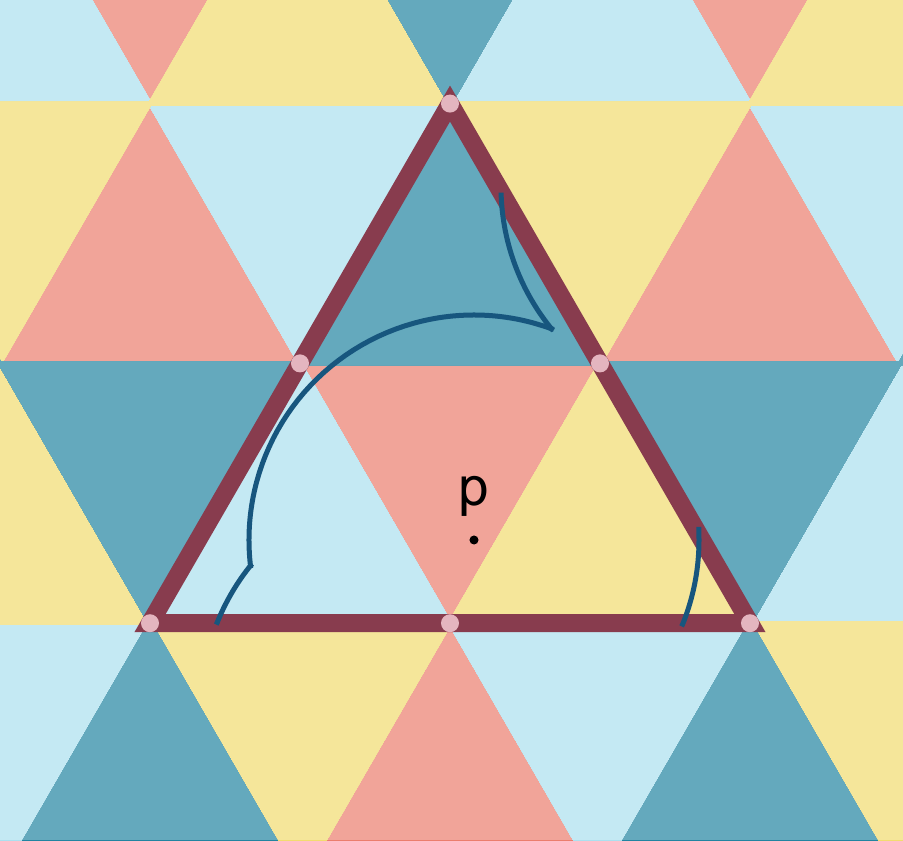}
\end{subfigure}
\caption{Constructing a geodesic circle on the tetrahedron (left) and mapping it inside the fundamental domain (right). }
\label{fig:wedge}
\end{figure}

Putting together the construction of geodesic circles $S_p^\delta$ and our earlier observation about the need to choose
the shortest geodesic between two points, we can now embark on computing average sphere distances for $\delta\!\in\! 
[0,1]$ in units of edge length. To obtain a pair of overlapping 
circles $S_p^\delta$, $S_{p'}^\delta$ for a given value of $\delta$, we start by picking a point $p$ 
randomly from the elementary region and constructing $S_p^\delta$, following the prescription of Sec.\ \ref{sec:geo}. 
In general, this will be given by a set of arcs in ${\cal C}_p$, parametrized by corresponding angle intervals in terms of the 
rotation angle $\theta$.
To simplify subsequent computations, we map all arc sections that 
happen to lie in a neighbouring domain ${\cal F}_i$ back to the fundamental domain $\cal F$ by appropriate
reflections and translations, which are symmetries of the tessellated plane (Fig.\ \ref{fig:wedge}). 

Next, we randomly pick a point $p'$ on $S_p^\delta$, which will serve as the centre of the second circle $S_{p'}^\delta$. We then again construct the arcs forming $S_{p'}^\delta$ using a region similar to ${\cal C}_p$, where the center point $p'$ is now generically not in the triangular elementary region. The analogous region ${\cal C}_{p'}$ can be found by an appropriate symmetry transformation. In principle it would be possible to then map all arc sections of $S_{p'}^\delta$ back into $\cal F$, but for the purpose of finding the minimum distance between $S_{p}^\delta$ and $S_{p'}^\delta$ it is simpler to perform two subsequent reflections of the arc sections through all six vertices on the boundary of $\cal F$. This puts 36 representatives of arcs of $S_{p'}^\delta$ in a neighborhood around $\cal F$, where some of these representatives can be discarded since several distinct pairs of subsequent reflections produce the same arc. The minimum distance from $S_p^\delta$ to any point $q' \in S_{p'}^\delta$ is guaranteed to be found among the remaining unique representatives.
Finding the average sphere distance $\bar{d}(S_p^{\delta},S_{p'}^{\delta})$ of eq.\ (\ref{sdist}) can now be done by 
integrating the Euclidean distance between all pairs of points $(q,q')\!\in\! S_p^{\delta}\times S_{p'}^{\delta}$.

\begin{figure}
\centering
\begin{subfigure}[t]{0.45\textwidth}
\includegraphics[height=5cm]{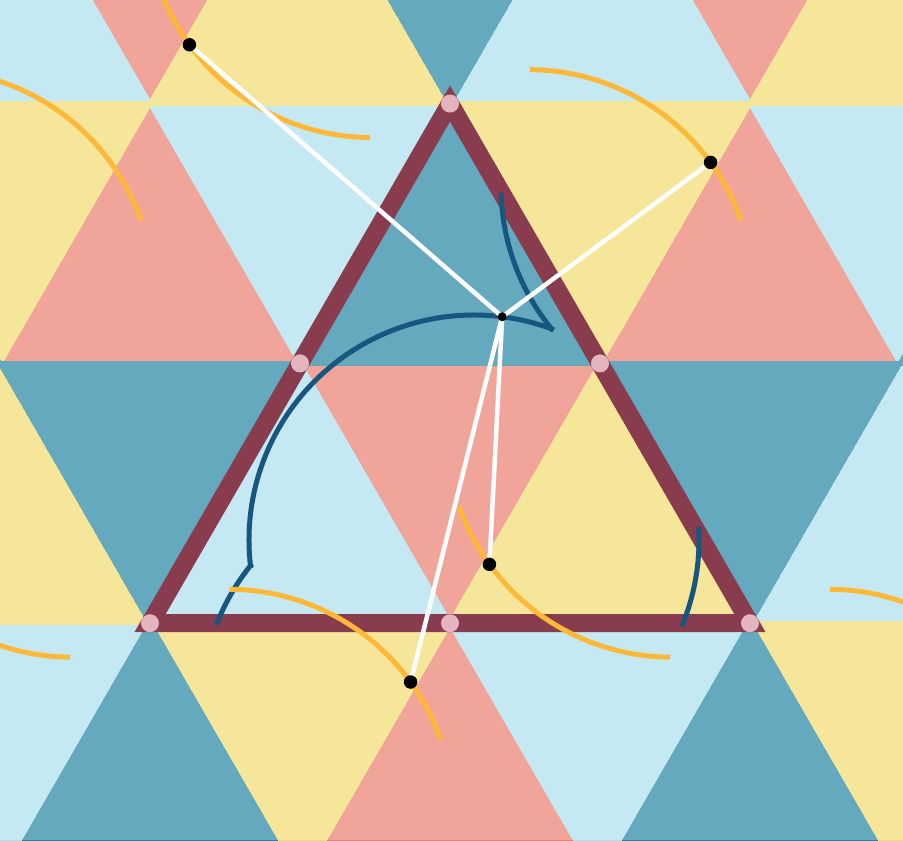}
\end{subfigure}
\begin{subfigure}[t]{0.45\textwidth}
\includegraphics[height=5cm]{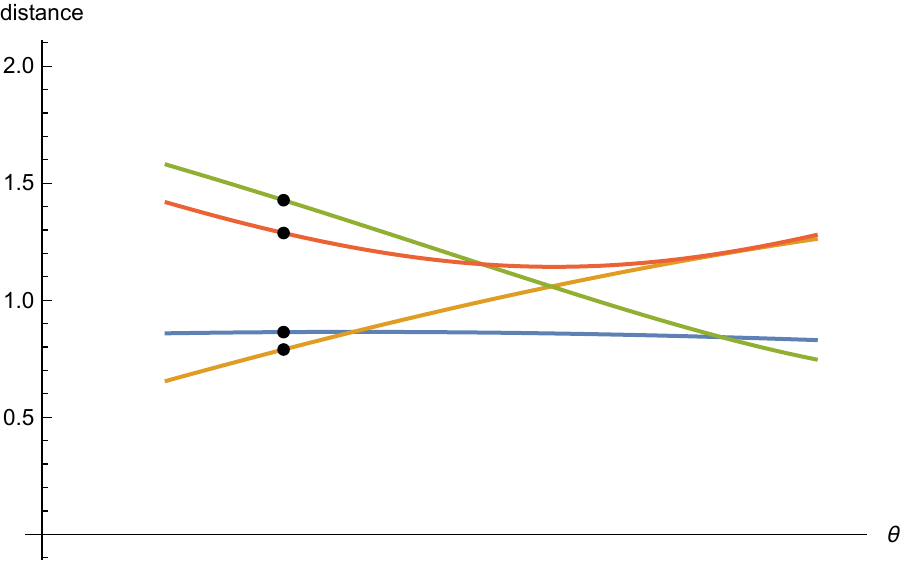}
\end{subfigure}
\caption{Computing the distance in the plane of a point $q \in S_{p}^\delta$ to an arc $q'(\theta)$ 
of the circle $S_{p'}^\delta$ in $\cal F$, as well as to several
copies $q_i'(\theta)$ of the arc in a neighbourhood of $\cal F$. Each coloured curve segment of distance measurements
in the graph on the right corresponds to a different reflection of the arc through one of the vertices. }
\label{fig:tetra-point}
\end{figure}

We show first how to measure the distance between a fixed point $q\! \in\! S_p^{\delta}$ and the second circle 
$S_{p'}^{\delta}$. The latter consists of a set of smooth arcs $q'(\theta')$, not necessarily contained in $\cal F$.
For sufficiently small $\delta$, and when $S_{p'}^{\delta}$ does not
enclose any singularities, there may be just a single ``arc", given by an entire smooth circle and parametrized by $\theta'\! \in\! [0,2\pi]$.
In general there will be several arcs, parametrized by smaller angle intervals. Whatever the case may be,
we determine the distance 
between $q$ and each such arc separately as follows. Consider a specific arc parametrized by $q'(\theta')\! \in\! [\theta'_{i},\theta'_{f}]$.
It is straightforward to
determine its distance from $q$ as a function of $\theta'$, using the Euclidean distance in the plane. However, recall from
our considerations at the beginning of Sec.\ \ref{sec:geo} 
that the corresponding geodesic (straight line) between $q$ and any specific point $q'(\theta')$ along the arc 
need not be the shortest one between the corresponding points on the tetrahedron. 
To find the shortest geodesic in a systematic way, we repeat the distance measurements with
each copy (representative) $q'_i(\theta')$ of the arc in the neighbourhood around $\cal F$. 

Fig.\ \ref{fig:tetra-point} shows an example where we have collected the distance measurements from four copies of an arc and
plotted the corresponding curve segments as a function of $\theta'\! \in\! [\theta'_{i},\theta'_{f}]$. 
The next step is to determine the continuous curve segment that to each $\theta'$ in this interval
assigns the minimum distance from $q$ to one of the four points $\{ q_1'(\theta'),\dots, q_4'(\theta') \}$.
This may be a single curve segment, which throughout the $\theta'$-interval lies below all 
other curve segments. Alternatively, it may consist of contributions from several mutually intersecting segments. 

The same construction must be applied to the remaining smooth arcs along $S_{p'}^{\delta}$. Putting the individual 
minimizing curve segments together results in a single continuous distance-minimizing curve. Integrating it over 
 $S_{p'}^{\delta}$, and dividing it by the length of $S_{p'}^{\delta}$ gives the average distance of $q$ to $S_{p'}^{\delta}$.  

To complete the computation of the average sphere distance (\ref{sdist}), we still need to vary the point $q$ 
over the arcs of the first circle, $S_{p}^{\delta}$. The analysis mirrors the one we just performed for the second circle.
It produces a two-dimensional version of the distance-minimizing curve, namely, a distance-minimizing ``sheet"
parametrized by two angles $(\theta,\theta')$,
obtained by computing and comparing distances $d(q(\theta),q'(\theta'))$ between all pairs of points from each pair of arcs from
$S_{p}^{\delta}$ and $S_{p'}^{\delta}$ respectively, and their representatives in neighbouring domains. These contributions must be combined, integrated and normalized to yield a single data point $\bar{d}(S_p^{\delta},S_{p'}^{\delta})$ 
at radius $\delta$. 

\begin{figure}[t]
\centering
\includegraphics[width=0.6\linewidth]{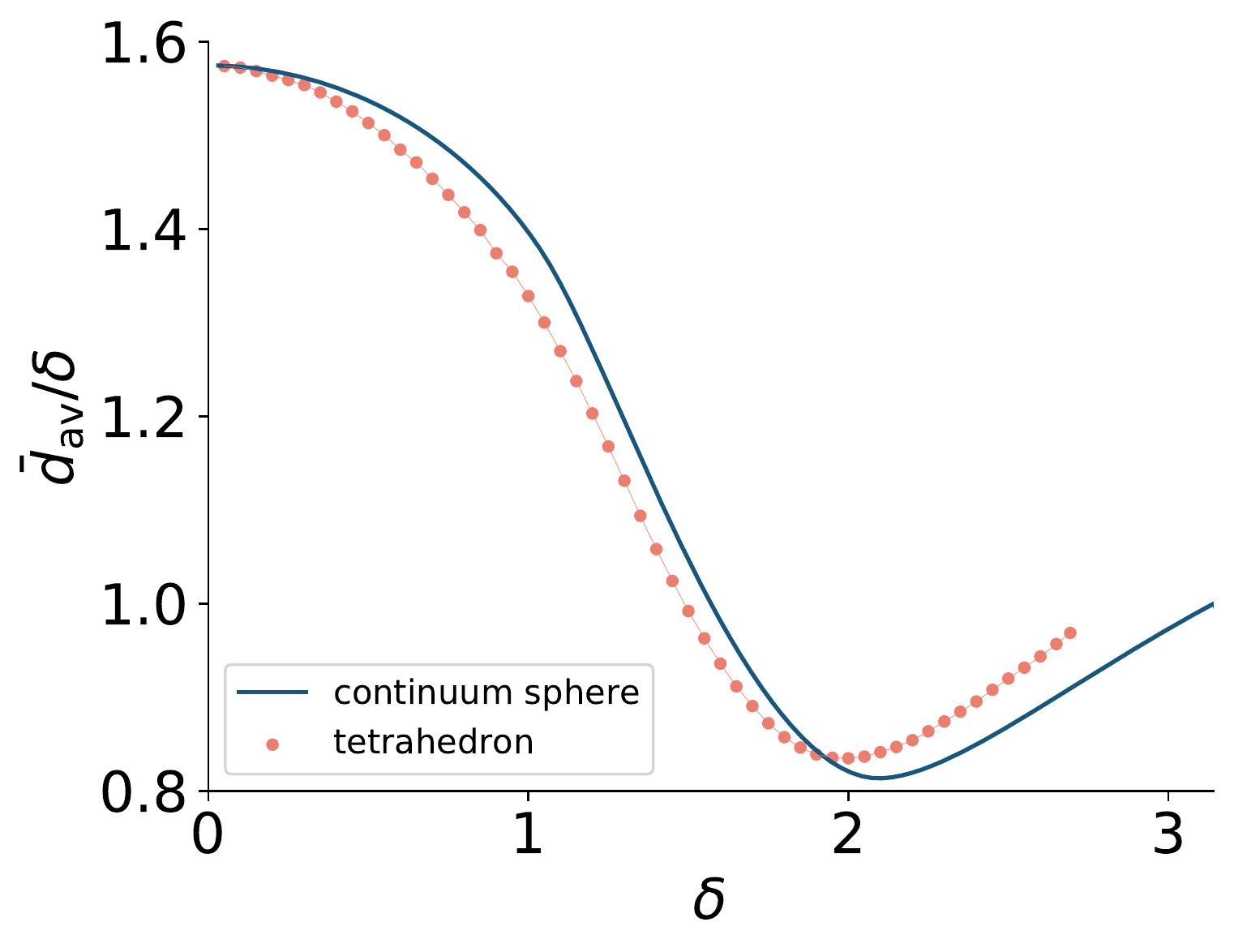}
\caption{Measurement of the curvature profile $\bar{d}_{\rm av}(\delta)/\delta$ 
of the surface of a tetrahedron, compared to that of a sphere of the same
area. (Error bars too small to be shown.) }
\label{fig:tetra-all}
\end{figure}

To collect the data for the curvature profile, we follow the same procedure as in Sec.\ \ref{sec:measure} to
generate double-sphere configurations that are distributed uniformly at random over the tetrahedron. 
We took random samples of 10.000 average sphere distance measurements at each of 54 evenly spaced
values of $\delta$. For easier comparison with the results obtained for the Platonic solids with our earlier 
method (Fig.\ \ref{fig:platonic}), we have reverted to ``volume-normalized" units where the edge length of the tetrahedron
is given by $L\! =\! 2.694$ (cf.\ Table 1). In these units, the step size for measurements is $\delta\! =\! 0.05$. 
The final result for the curvature profile $\bar{d}_{\rm av}(\delta)/\delta$ is shown in Fig.\ \ref{fig:tetra-all}, with the curvature profile of
the smooth two-sphere for comparison. 
For small $\delta$, the data are compatible with those obtained with our earlier method. 

Overall, the curvature profile of the tetrahedron is qualitatively similar to that of the sphere, but clearly distinct from it,
in a way that cannot be absorbed by a simple linear rescaling of the $\delta$-axis.
The similarity demonstrates in explicit terms the robustness of this global observable, by which we mean its property of ``averaging out" a
rather extreme curvature distribution, like that of the tetrahedron. It strengthens earlier observations of such a behaviour on
large Delaunay triangulations of a sphere \cite{qrc1}. These are also examples of piecewise flat spaces, but with a curvature distribution
that is much closer to that of a continuum sphere, with small deficit angles everywhere. 

As the measurements of Sec.\ \ref{sec:measure} already indicated, the curve for $\bar{d}_{\rm av}/\delta$ falls more steeply for
small $\delta$,
which means that the (averaged) quasi-local quantum Ricci curvature of the tetrahedron is larger than that of the sphere.  
Features of the sphere curve for larger values $\delta\! \gtrsim \! 0.5$ account for large-scale geometric and topological properties
of the sphere and are reflected in a similar behaviour of the tetrahedron, although the minimum of the curve is reached for 
smaller $\delta$ and its value is larger. 
When interpreting the large-scale behaviour, one should keep in mind that on the two-sphere the circles $S_p^\delta$ have maximal size for
$\delta\! =\! \pi/2$, after which they start shrinking until for $\delta\! =\! \pi$ they degenerate into points. 
The latter is reflected in the value $\bar{d}_{\rm av}(\pi)/\pi\! =\! 1$ on the sphere, the endpoint of the curve in Fig.\ \ref{fig:tetra-all}. 
The fact that each point $p$ on the sphere has
an antipode at geodesic distance $\pi$ is of course a consequence of the highly symmetric character of its geometry. As we
have already seen, this is different on the tetrahedron, where the distance of the antipode depends on the point $p$ and there is
no analogue of the distinguished value $\delta\! =\! \pi$. 
The restriction on the $\delta$-range we set for the tetrahedron was precisely the distance of the closest antipode. 
We could in principle have chosen to go beyond this point, assigning the value zero to data points whose antipodal distance
is smaller than a given $\delta$. This would have extended the tetrahedron's curve to $\delta\! \approx \! 3.110$, but at the expense of
a somewhat unclear interpretation. 

\section{Summary and conclusion}
\label{sec:final}

Motivated by the issue of observables in nonperturbative quantum gravity, 
we have advocated the study of a new, global geometric observable for curved metric spaces, the curvature profile.
It is obtained by integrating the quasi-local, scale-dependent quantum Ricci curvature introduced in earlier work, 
and has the interpretation of an averaged Ricci scalar depending on a scale $\delta$. On smooth classical
manifolds, the information contained in the curvature profile for infinitesimal and small $\delta$ is simply that of the 
averaged Ricci scalar, while for larger $\delta$ it captures the geometric and topological properties of the metric space 
in a coarse-grained manner. This scale dependence is very important for the corresponding observable in the quantum
theory, where it can help us to identify the transition from a pure quantum regime to a semiclassical one. In order
to be able to identify the latter, it is important to get a better understanding of the behaviour of
the curvature profile on classical spaces.    

In the present work, we have specifically examined the influence of the distribution of the local curvature of the underlying
metric space on the classical curvature profile. The easiest set-up to compute and compare this effect explicitly is that of 
two-dimensional compact surfaces. We have investigated several regular polyhedral surfaces homeomorphic to the two-sphere, which 
contain a number of conical singularities. We were able to analyze the case of the tetrahedron completely, because we 
could set up a relatively simple method to compute the geodesic distance between two given points, making use
of a representation of the geodesics in the tessellated plane.\footnote{A related problem for the more difficult case of the
cube has been addressed in \cite{shortest}.} The curvature profile is distinct from that of the sphere, although its overall
features are similar. Perhaps the most remarkable feature is how well it resembles the profile of a smooth sphere. 
The property of averaging or coarse-graining well over regions where the curvature is singular is a
desirable feature from the point of view of the nonperturbative quantum theory, where the quantum geometry 
on Planckian scales tends to be extremely singular and ill-defined locally. Of course, the quantum Ricci curvature
underlying the construction of the curvature profile was introduced precisely to address and potentially mitigate this issue. 
For the other Platonic solids we could determine the curvature profile only for small $\delta$, but from the
limited evidence it appears that distributing the Gaussian curvature over more vertices leads to profiles that are even closer to
that of the sphere. 

In conclusion, we have for the first time computed a curvature profile for a curved classical space that is not maximally 
homogeneous and isotropic. It gives us a first quantitative gauge of how deviations from a maximally symmetric
situation are reflected in the curvature profile, which constitutes a kind of global fingerprint of a given geometry. 
It will be interesting to add the curvature profiles of other classical geometries to our reference catalogue, to help
us bridge the divide between results from nonperturbative quantum gravity and invariant properties of classical spacetimes.
\vspace{0.5cm}

\noindent {\bf Acknowledgments.} 
This work was partly supported by a Projectruimte grant of the Foundation for Fundamental Research 
on Matter (FOM, now defunct), financially supported by the Netherlands Organisation for Scientific Research (NWO).


\vspace{0.3cm}


\begin{thebibliography}{99}

\bibitem{livrev}
R.\ Loll, 
{\it Discrete approaches to quantum gravity in four dimensions}, 
Living Rev.\ Rel.\ 1 (1998) 13 [arXiv:gr-qc/9805049].

\bibitem{review1}
J.~Ambj\o rn, A.~G\"orlich, J.~Jurkiewicz and R.~Loll,
{\it Nonperturbative quantum gravity},
Phys.\ Rep.\  519 (2012) 127-210 [arXiv:1203.3591, hep-th].

\bibitem{review2}
R.~Loll,
{\it Quantum gravity from Causal Dynamical Triangulations: A review},
Class.\ Quant.\ Grav.\ 37 (2020) 013002 [arXiv:1905.08669, hep-th].

\bibitem{toruscoord}
J.~Ambj\o{}rn, Z.~Drogosz, J.~Gizbert-Studnicki, A.~G\"orlich and J.~Jurkiewicz,
{\it Pseudo-Cartesian coordinates in a model of Causal Dynamical Triangulations},
Nucl.\ Phys.\ B (2019) 114626 [arXiv:1812.10671, hep-th].

\bibitem{CDT1} 
J.\ Ambj\o rn, J.\ Jurkiewicz and R.\ Loll, 
{\it Emergence of a 4D world from causal quantum gravity}, Phys.\ Rev.\ Lett.\ 93 (2004) 131301 [arXiv: hep-th/0404156].

\bibitem{CDT2} 
J.\ Ambj\o rn, J.\ Jurkiewicz and R.\ Loll, 
{\it Semiclassical universe from first principles}, 
Phys.\ Lett.\ B\ 607 (2005) 205-213 [arXiv:hep-th/0411152].

\bibitem{spectral}
J. Ambj\o rn, J. Jurkiewicz and R. Loll, 
{\it Spectral dimension of the universe},
Phys.\ Rev.\ Lett.\ 95 (2005) 171301 [arXiv:hep-th/0505113].

\bibitem{reconstructing}
J. Ambj\o rn, J. Jurkiewicz and R. Loll, 
{\it Reconstructing the universe}, 
Phys.\ Rev.\ D\ 72 (2005) 064014 [arXiv:hep-th/0505154].

\bibitem{desitter1}
J.~Ambj\o rn, A.~G\"orlich, J.~Jurkiewicz and R.~Loll,
{\it Planckian birth of the quantum de Sitter universe},
Phys.\ Rev.\ Lett.\ 100 (2008) 091304 [arXiv:0712.2485, hep-th].

\bibitem{desitter2}
J.~Ambj\o rn, A.~G\"orlich, J.~Jurkiewicz and R.~Loll,
{\it The nonperturbative quantum de Sitter universe},
Phys.\ Rev.\ D\ 78 (2008) 063544 [arXiv:0807.4481, hep-th].


\bibitem{qrc1}
N.\ Klitgaard and R.\ Loll, 
{\it Introducing quantum Ricci curvature},
Phys.\ Rev.\ D\ 97 (2018) 046008 [arXiv:1712.08847, hep-th].

\bibitem{qrc2}
N.\ Klitgaard and R.\ Loll, 
{\it Implementing quantum Ricci curvature}, 
Phys.\ Rev.\ D\ 97 (2018) 106017 [arXiv:1802.10524, hep-th].

\bibitem{qrc3}
N.\ Klitgaard and R.\ Loll, 
{\it How round is the quantum de Sitter universe?}, 
Eur.\ Phys.\ J.\ C 80 (2020) 10, 990 [arXiv:2006.06263, hep-th].

\bibitem{ollivier}
Y.\ Ollivier,
{\it Ricci curvature of Markov chains on metric spaces},
J.\ Funct.\ Anal.\ 256 (2009) 810-864.

\bibitem{tru1}
C.A.~Trugenberger,
{\it Random holographic ``large worlds" with emergent dimensions},
Phys.\ Rev.\ E\ 94 (2016) no.5, 052305 [arXiv:1610.05339, cond-mat.stat-mech].

\bibitem{tru2}
C.A.~Trugenberger,
{\it Combinatorial quantum gravity: geometry from random bits},
JHEP\ 1709 (2017) 045 [arXiv:1610.05934, hep-th].

\bibitem{tru3}
C.\ Kelly, C.A.\ Trugenberger and F.\ Biancalana,
{\it Self-assembly of geometric space from random graphs},
Class.\ Quant.\ Grav.\ 36 (2019) 125012 [arXiv:1901.09870, gr-qc].

\bibitem{tru4}
C.~Kelly, C.~Trugenberger and F.~Biancalana,
{\it Emergence of the circle in a statistical model of random cubic graphs}
[arXiv:2008.11779, hep-th].

\bibitem{fuchs1}
D.\ Fuchs and E.\ Fuchs,
{\it Closed geodesics on regular polyhedra},
Mosc.\ Math.\ J.\ 7 (2007) 265-279.

\bibitem{fuchs2}
D.\ Fuchs, 
{\it Periodic billiard trajectories in regular polygons and closed geodesics on regular polyhedra}, 
Geom.\ Dedicata 170 (2014) 319?333.

\bibitem{shortest}
R.\ Goldstone, R.\ Roca and R.\ Suzzi Valli,
{\it Shortest paths on cubes} [arXiv:2003.06096, math.HO]


\end{thebibliography}
\end{document}